\documentclass[prc,aps,12pt,final,notitlepage,oneside,twocolumn,nobibnotes,nofootinbib,superscriptaddress,noshowpacs]{revtex4}
\usepackage{amssymb,amsmath}
\usepackage{mathtext}
\usepackage{graphicx,colordvi}
\usepackage{color}
\usepackage{calc}
\usepackage{ifthen}

\def\be{\begin{equation}}
\def\ee{\end{equation}}
\def\bea{\begin{eqnarray}}
\def\eea{\end{eqnarray}}
\def\l{\label}

\def\hahat{\hat{H}}

\def\hahat0{\hat{H}_0}

\def\sin{\hbox{sin}}

\def\exp{\hbox{exp}}

\def\d{\hbox{d}}
\def\eps{\varepsilon}
\def\epsi{\mathcal{E}}

\def\siml{\hbox{\kern.1em \lower.6ex \hbox{$\sim$} \kern-1.12em
 \raise.6ex \hbox{$<$} \kern.1em}}
\def\simg{\hbox{\kern.1em \lower.6ex \hbox{$\sim$} \kern-1.12em
 \raise.6ex \hbox{$>$} \kern.1em}}
\begin{document}

\title{
  LEPTODERMIC CORRECTIONS TO THE TOV EQUATIONS AND
  NUCLEAR ASTROPHYSICS
  WITHIN THE EFFECTIVE SURFACE
 APPROXIMATION}
\author{A.G. Magner}
\affiliation{\it  Institute for Nuclear Research, 03028 Kyiv, Ukraine}
\affiliation{\it Cyclotron Institute, Texas A\&M University, College Station, Texas 77843, USA}
\author{S.P.~Maydanyuk}
\email{sergei.maydanyuk@impcas.ac.cn}
\affiliation{\it  Institute for Nuclear Research, 03028 Kyiv, Ukraine}
\affiliation{\it Southern Center for Nuclear-Science Theory (SCNT), Institute of Modern Physics, Chinese Academy of Sciences, Huizhou 516000, China}
\author{A.~Bonasera}
 \affiliation{\it Cyclotron Institute, Texas A\&M University,
   College Station, Texas 77843, USA}
 \author{H.~Zheng}
 \affiliation{\it School of Physics and Information Technology, Shaanxi Normal University,Xi'an 710119, China}
 \author{T.~Depastas}
 \affiliation{\it Cyclotron Institute, Texas A\&M University,
  College Station, Texas 77843, USA}
\author{A.I. Levon}
\affiliation{\it  Institute for Nuclear Research, 03028 Kyiv, Ukraine}
\author{U.V.~Grygoriev }
\affiliation{\it  Institute for Nuclear Research, 03028 Kyiv, Ukraine}

\date{\today}

\begin{abstract}
  The macroscopic model for a neutron star (NS) as 
 a liquid drop at
  the equilibrium
  is used to extend the Tolman-Oppenheimer-Volkoff (TOV) equations
  taking into account the gradient terms
  responsible for the system surface. The parameters of the Schwarzschild
  metric in the spherical case are found with these surface corrections
  to the known leading (zero) order of
  the leptodermic approximation 
    $a/R<<1$, where $a$ is the NS effective-surface (ES) thickness, and
 $R$ is the effective NS radius.
 The energy density $\epsi$ is considered in a general form including
 the functions of the particle number density and of its gradient terms.
    The macroscopic gravitational component $\Phi(\rho)$ of the energy density
  is taken into account  
    in the simplest form as expansion in powers of
   $\rho-\overline{\rho} $, where
        $\overline{\rho}$ is the saturation density,
    up to second order, in terms of its contributions to the
    separation particle energy and incompressibility. 
  Density distributions $\rho$ across the 
  NS ES in the normal direction to the ES, which are derived
  in the simple analytical
  form at the same leading approximation, was
  used for the derivation of
  the modified TOV (MTOV) equations by accounting for their
  NS surface corrections. The MTOV equations are analytically
  solved at first order and the results
  are compared with the standard
  TOV approach of the zero order.

\end{abstract}

\maketitle


\section{INTRODUCTION}
\label{introd}

The macroscopic effective-surface approach (ESA)
\cite{wilets,strtyap,tyapin,strmagbr,strmagden,MS09,BM13,BM15}
taking into account the gradient terms of the
energy density was applied \cite{MM24} for the neutron stars.
Following the works of Tolman,
Oppenheimer, and Volkoff (TOV) \cite{RT39,OV39} we consider the neutron star
(NS) as
a finite, dense and perfect liquid drop at
equilibrium under the gravitational,
and other realistic  forces;
see also the Tolman book \cite{RT87} (chapt. 7, sect. 96).
As shown in Ref.~\cite{RT87},
the TOV equations were derived \cite{OV39} from the
    equations of the
General
Relativistic Theory (GRT) assuming the spherical symmetry and the
macroscopic approximation for the energy-momentum tensor.
The expression for the
energy-momentum tensor was essentially used in the
TOV derivations  for 
    a macroscopic system as
the perfect
liquid drop (see Refs.~\cite{LLv2} (chapts. 11 and 12),
\cite{LLv6} (chapt.~1, sects.~1-3; chapt.~7, sect.~61;
chapt.~15, sect.~133) and Rowlinson-Widow \cite{RW82} (chapt.~1).
Many applications  of the TOV result
    in terms of the Schwarzschild metric 
    solutions, Ref.~\cite{KS16},
for spherically symmetric NSs  can be
found in
 Refs. \cite{LLv2,WF88,CB97,ST04,SH06,HPY07,Ko08,PFCPS13,AB14,BC15,LH19,SBL23}
(sect.~9 in Ref.~\cite{ST04} and sect.~3 in Ref.~\cite{HPY07}).
   For the basic relations of the nuclear and neutron liquid-drop
    models
    to NS properties
    we should also mention a pioneer work \cite{BBP71}
   done by
Baym, Bethe, and Pethick. They discussed the
liquid matter drop with the leptodermic property
having a sharp decrease of the particle
number density $\rho$
in a relatively small edge region considered as its surface.
Notice, in the TOV derivations, the authors assumed that
    the outer and inner Schwarzschild metrics have to be
    matched at a definite NS radius that looks meaningful for a leptodermic
    behavior of the density $\rho$.
    For a good hint to the nuclear astrophysics; see also, e.g.,
        Refs.~\cite{LP01,HP01PRL,HP01PRC,PH00Poland,CH08,ABCG09,FCPG13,LH19,FG23,DFG23,LJ23,Pe23,Pe24,XV24}.
   Clear specific definitions and complete updated results
    for the energy density with
density gradient (surface) terms and  for equations of
the infinite matter state
 with many inter-particle
forces in the non-relativistic and relativistic cases
can be found in the 
recent review Ref.~\cite{SBL23}.

\begin{figure}
  \vskip1mm
    \includegraphics[width=8.0cm]{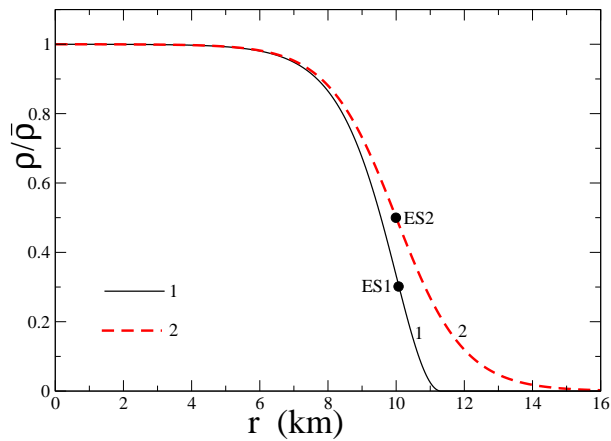}

    \vspace{-0.5cm}
  \caption{{\small
      Particle number density, $\rho$, in units of the saturation value,
      $\overline{\rho}$,
          as function of the radial variable $r$ (in km)
          for a 
          neutron star
      in a simple 
      compressed
      liquid-drop model at the stable
      equilibrium.
            Solid line is related to the asymmetric solution,
             Eq.~(32)$^\ast$ [the asterisk
              means the equation taken from
                  Table \ref{table-1} of Ref.~\cite{MM24}, 
      line ``1''],
      which is the leading approximation over a small
      leptodermic parameter, $a/R\ll 1$.
      Dashed line presents the same but for the Wilets symmetric solution
      \cite{wilets,MM24}
         (line ``2'').
      Parameters, for example:
         the effective radius $R=10$ km, diffuseness of the NS crust
          $a=1$ km, 
       and temperature
      $T=0$;  see, e.g., Refs.~\cite{ST04,HPY07,MM24}.
         The dots ES1 and ES2 show the ES at the effective radius $R$ for
         the solid
                    and dashed
         lines, respectively (from Ref.~\cite{MM24}).
}}
\label{fig1}
\end{figure}

Using the step-function for the particle number density $\rho$, 
    which is a
    constant inside and zero outside of the system, in line of the
    Schwarzschild
    derivations \cite{KS16},
 Tolman obtained \cite{RT87}
 the explicit analytical solutions for the pressure.
    They were found in terms of the
 Schwarzschild metric
 coefficients for the spherical system symmetry, and the
 macroscopic energy-momentum tensor for
 the perfect
 liquid drop. Analytical expressions for the pressure in terms of the
 simple one-dimensional integrals can be obtained
 also for the leptodermic
 particle-number density $\rho$, e.g., as shown in Fig.~\ref{fig1}.

\begin{table*}[pt]
\begin{tabular}{|c|c|c|}
\hline
Equations & Symbols & Meaning
\\
\hline
(3) & $\epsi$ & Energy density\\
\hline
 (8) & $\Phi$ & Macroscopic gravitational field\\
\hline
(9) & $\mathcal{A}$ & Volume energy density\\
\hline
(10) & $\varepsilon^{}_G$ &
 Volume energy density component\\
\hline
(11) & $K_G$ & Total incompressibility \\
\hline
(12) & $\mathcal{B}$ & Surface energy density coefficient\\
\hline
(16) & $\mathcal{A}_V$ & Volume energy density\\
\hline
(19)  & $\epsilon$ & Dimensionless energy density\\
\hline
(21) 1st & $\rho=\rho^{}_V$ & Volume particle number density\\
\hline
(21) 2nd & $\mathcal{M}$ & Chemical potential correction\\
\hline
(24) & Zero order equation & For particle density $\rho$\\
\hline
(27) & Zero order equation & For dimensionless density $y(x)$\\
\hline
(31) & $x(y)$ & Leading order solution \\
\hline
(32) & $y(x)$ &
vdW-Skyrme solution \\
\hline
(60) & $P$ &
Total pressure \\
\hline
(61) & $P^{}_{\rm V}$ &
Volume pressure \\
\hline
(62) & $P_{\rm S}$ &
Surface pressure \\
\hline
\end{tabular}

\vspace{-0.2cm}
\caption{{\small
    Equations of Ref.~\cite{MM24} (the first column),
    marked by the asterisk
    in the text, are presented for the quantity 
    with symbols in the second column, and explained in the third
    column. The vdW-Skyrme is the van der Waals and Skyrme inter-particle
    interaction.
  }}
\label{table-1}
\end{table*}
Notice that the Schwarzschild radius $R_S$ \cite{RT39,OV39,RT87}
    can be considered as
    a constant which determines the upper limit of the NS radius 
        and can be
    used for the transition to the black holes if the particle density $\rho$
    has a saturation property $\rho \approx \overline{\rho}$ inside of the
    NS. Therefore, the  energy density $\epsi$ which is
    the argument of $R_S$, $R_S\propto (\epsi(\overline{\rho}))^{-1/2}$,
    has also the corresponding saturation property. Otherwise, we have to
    modify the TOV equations for the pressure $P(r)$ to agree with the
    equation of state (EoS)
    which would allow a smooth behavior of the density $\rho(r)$.
     As the TOV equation \cite{OV39} is used along with
    the EoS in a lot
    of astrophysical articles on the NS,
    it would be logical to
    agree the arguments for the
    specific derivations of both
    these equations. This is the main motivation of our and TOV
    macroscopic approaches. Notice also that of course, in nuclear astrophysics 
    we have to take into account
    the strong gravitational forces as in 
        the derivations
    of the TOV equations,
    in contrast to the nuclear physics. Our suggestion for these coordinations
    does not exclude more microscopic EoSs. However, it requires 
    the corresponding
    modification of the TOV equations 
        which, unfortunately, makes the problem
    to be more complicate, far away from the present analytical approach.

In this work we 
derive the
 modified TOV (MTOV) equations by taking into
 account the gradient surface 
 components of the energy density  which are suggested basically, e.g., in
     Refs.~\cite{BBP71,LP01,HP01PRL,HP01PRC,PH00Poland,HPY07,CH08,ABCG09,PFCPS13,FCPG13,LH19,FG23,DFG23,SBL23,LJ23,Pe23,Pe24,XV24,MM24}.
Within the leptodermic  approximation \cite{MM24} based on such
     gradient density components of the energy density, 
         one can consider
     first analytically a dense one-component system. 
     In these calculations we take the energy density functional $\epsi(\rho)$
 in a rather
 general form,  as linear in
     $(\nabla\rho)^2$ terms, with the coefficients
 which are smooth functions
 of the particle density $\rho$.
 In such a system, the particle density $\rho$ in the leading order
 approximation is function of the
 radial coordinate with
exponentially decreasing behavior 
from almost a constant saturation value inside of the dense system
to that of zero through the NS effective
surface in
a relatively small crust range $a$; see, e.g., Fig.~\ref{fig1}.
The ES is defined 
as the inflection point of spatial coordinates, i.e.,
with density gradients maximum.
In order to obtain the analytical solutions for the
particle number density and EoS, we 
use the
 leptodermic macroscopic
 approximation.
 In this approximation,
the gravitational forces
can be taken into account through the macroscopic gravitational
field $\Phi$ determined by 
    the second order expansion over deflection
of the density $\rho$ from its saturation value
$\overline{\rho}$
(see 
Ref. \cite{MM24}).

Within the leptodermic approximation, $a/R \ll 1$,
simple and
accurate solutions of many nuclear and liquid-drop
problems involving the particle number
density distributions 
were obtained by using the ES approach
for nuclei
\cite{wilets,strtyap,tyapin,strmagbr,strmagden,MS09,BM13,BM15};
see also Ref.~\cite{MM24} for applications to the neutron stars.
The ESA exploits
the property of saturation of the nuclear particle density $\rho$
inside of the system,
which is a characteristic
feature of dense systems
such as molecular systems \cite{RW82,vdW},
liquid drops (amorphous solids),
nuclei,
 and presumably, NSs, in agreement with the TOV derivations. 
 The realistic energy-density distribution per particle
 is minimal at a certain saturation
density $\overline{\rho}$,  
corresponding approximately to the infinite dense matter \cite{bete}.
As a result, 
relatively narrow edge region exists in finite nuclei or NS (crust)
in which the 
density drops sharply from its almost central value to zero. We assume
here that the part 
inside of the system far from the ES can be  changed
a little (saturation property of the dense system 
  as  hydrostatic liquid drop, nucleus,
or presumably, NS in a final evolution state).

The coordinate system related to the effective surface
is defined in such 
a way that one of the spatial coordinates ($\xi$)
is the distance from a 
given point to the 
ES, and others ($\eta$) are tangent to the ES.
 This nonlinear coordinate system $\xi,\eta$ is
conveniently used in 
a region near the nuclear and NS edge.
They  allow for an easy extraction of the
relatively large 
terms in 
density distribution equations 
    for the condition of a
variational 
    equilibrium.
      This condition means that 
      the variation of
      the total energy $E$ over the
density $\rho$ is zero under the constraints which fix 
some
integrals of
motion beyond the energy $E$ by the Lagrange method. The Lagrange multipliers
  are determined by these constraints
  within the local energy-density theory, in particular,
  the extended Thomas-Fermi 
  (ETF) approach 
  from 
      nuclear, dense molecular, and metallic-cluster
  physics \cite{brguehak,brbhad}.
  Neglecting the other 
      smaller-order perpendicular- and all parallel-to-ES
  contributions, 
sum of such terms leads to a simple one-dimensional equation (in special local 
coordinates with the coordinate normal-to-surface $\xi$).
Such an equation mainly determines approximately the density
distribution 
across the diffused surface layer. The latter is of the relatively
small order over
the ratio of the diffuseness 
    parameter $a$ 
to the (mean curvature) ES radius $R$,
$a/R << 1$, for sufficiently
heavy systems. 
A small parameter, $a/R$, of the expansion within the ES approach can be used
for analytical solving the variational problem
for the minimum of the energy density functional per 
volume with constraints for
a fixed particle number, and other integrals of motion,
 such as angular momentum,
quadrupole deformation, etc.   
When this edge distribution of the 
density is known, the leading
static and dynamic density distributions which 
correspond to the
diffused surface conditions can be easily constructed. 
To do so, one has 
to determine the dynamics of the effective 
surface which is coupled to 
    the volume dynamics of the density by certain 
liquid-drop model 
boundary conditions \cite{bormot,strmagbr,magstr}.
 Our ESA
is
based on the catastrophe theory for solving differential equations with
a small parameter of the order of $a/R$ as the coefficient
in front of the high order derivatives in normal to the ES direction
 \cite{Ti52}.
     A relatively large change of the density $\rho$ on a small
     distance $a$ with respect to the curvature radius $R$ takes place for
     a finite
 liquid-matter drop
 (nuclei, molecular water drops, neutron stars).
 We emphasize
 that the macroscopic ESA theory \cite{MM24}
 allows to consider the deformed and super-deformed systems for which the
 same coordinate system $\xi,\eta$ can be used locally near the ES. 
Inside of such dense systems,
the density $\rho$ is changed a
little around
constant saturation density $\overline{\rho}$. Therefore, including also
the second order expansion of the gravitational
field near $\overline{\rho}$,
one obtains
essential effects
of the surface capillary pressure 
    which were originally studied for
molecular systems by van der Waals; see
Refs.~\cite{RW82,vdW}.

The accuracy of the ESA was checked in 
Ref.~\cite{strmagden} for the case of 
    static and dynamic nuclear physics
by comparing the results 
with the existing nuclear theories like Hartree-Fock (HF)
\cite{vauthbrink} and ETF \cite{brguehak,brbhad}, based on the 
Skyrme forces 
\cite{vauthbrink,skyrme,barjac,ringshuk,blaizot,brguehak,gramvoros,krivin,CB97,CB98}, 
but for the simplest
case without spin-orbit and asymmetry terms of the energy density functional.
The direct variational principle for finding numerically 
the parameters of the tested particle number
density functions in simple forms
of the Woods-Saxon-like in Ref. \cite{brbhad}
 or their powers
(Ref.~\cite{kolsan})
 were applied by using the realistic Skyrme energy 
     functional \cite{CB97,CB98};
see also recently published
variational ETF approach \cite{KS18,KS20}.
The extension of the ES approach
to the nuclear isotopic symmetry and spin-orbit interaction has been done in
 Refs. \cite{MS09,BM13,BM15}. The Swiatecki derivative
terms of the symmetry energy for
heavy nuclei
\cite{myswann69,myswiat85,danielewicz2,vinas1,vinas2,vinas4,vinas5,Pi09}
were
taken into account  within the ESA in Ref. \cite{BM15}. The
discussions of the progress in nuclear physics and astrophysics
within the relativistic local density approach,
can be found in review articles,
Refs.~\cite{SBL23,NVR11}; see also
Refs.~\cite{CH08,Pi21,Pe23,Pe24}.

In the present work we 
apply the effective surface approach
for the neutron stars of Ref.~\cite{MM24},
based on Refs.~\cite{strmagden,MS09,BM13,BM15} of nuclear physics,
to derive the MTOV equations taking into account the surface leptodermic
corrections. 
In Sect.~\ref{tov} 
we present the TOV derivations at the zero order approach.
In Sect.~\ref{mtov} we derive the MTOV equations with 
    accounting for the
 first order surface corrections to the
TOV equations. 
 The derivations of
 the analytical solutions to the MTOV equations are
 obtained too. Section \ref{disc}
is devoted to the discussions of the results. These results are summarized in
the conclusion section \ref{concl}. Some details of the calculations
are given in
Appendix \ref{appA}.

\section{Tolman-Oppenheimer-Volkoff approach}
\l{tov}

Starting from the Einstein-Gilbert
equations of the GRT for the spherically symmetric
system, one has the
Schwarzschild gravitational metric in a general form:
\be\l{ds2schw}
    {\rm d} s^2=-e^\lambda{\rm d} r^2-
     r^2 {\rm d}\theta^2 -
     r^2 \sin^2\theta {\rm d} \phi^2
      + c^2e^\nu {\rm d} t^2,
\ee
where $c$ is the speed of light.
Functions  $\lambda=\lambda(r)$ and $\nu=\nu(r)$ are given by the
differential equations (see Ref.~\cite{RT87} at zero
cosmological
constant $\Lambda$). First,  outside of the system on big distances
from masses, or for 
    relatively
small gravitational forces 
\cite{RT87,LLv2}, one can derive these equations 
using the macroscopic properties of a dense spherical
system for the energy-momentum
 tensor. 
Transforming then the radial coordinate $r$ of the
spherical four-coordinate system $r,\theta,\phi$, and $t$, 
due to the GRT in-variance, for the line element in this space,
Eq.~(\ref{ds2schw}),
one obtains the simple differential
equations for $\lambda$ and $\nu$ of the
Schwarzschild metric for the region inside of the liquid-drop system
(see, Refs.~\cite{RT87,LLv2}),
\bea\l{lamdanueq}
& \frac{8\pi G}{c^4} P
=e^{-\lambda}\left(\frac{\nu^\prime}{r} + \frac{1}{r^2}\right) -
\frac{1}{r^2}~,\nonumber\\
& \frac{8\pi G}{c^4} \epsi
=  e^{-\lambda}\left(\frac{\lambda^{\prime}}{r}
-\frac{1}{r^2}\right) +
\frac{1}{r^2}~,\nonumber\\
&\frac{d P}{d r}=-\frac{1}{2}\left(\epsi + P\right)\nu^\prime~,
\eea
where $G$ is the gravitational constant.
Primes above $\lambda$ and $\nu$ mean the corresponding derivatives over the
radial coordinate $r$, and $\epsi$ is 
    the energy density. 
With the EoS, $\epsi=\epsi(\rho)$
[or, $P=P(\rho)$], one arrives at the complete system of equations 
    for $\lambda$,
$\nu$, $\epsi$ (or density $\rho$), and $P$.

Using the substitution $e^{-\lambda}=f$,
one can solve the second equation in Eq.~(\ref{lamdanueq}),
as a linear first-order differential
    equation over $f(r)$, in terms
    of the quadratures. Then, the derivative
    $\nu^\prime$ can be found from the first
equation of Eq.~(\ref{lamdanueq}). Integrating it over $r$, one obtains the
Schwarzschild
metric (\ref{ds2schw}) in terms 
    of these solutions for $\lambda(r)$ and
$\nu(r)$.
Substituting
the derivative $\nu^\prime$ into the third equation of Eq.~(\ref{lamdanueq}),
one finally obtains the
famous TOV equations \cite{RT39,OV39,LJ23},
\bea\l{toveq}
&\frac{{\rm d} P}{{\rm d} r}= 
-\frac{G(\epsi + P )(mc^2 + 4\pi r^3P )}{rc^2(rc^2 - 2G m)} ,\nonumber\\
&\frac{{\rm d} m}{{\rm d} r}
= \frac{4\pi r^2\epsi}{c^2}.
\eea
Here, $m(r)$ is the gravitational mass interior to the radius $r$, 
    and
$\epsi$ is a given energy density.
     Boundary conditions for these equations are given by
     \be\l{boundcondmp}
     m(r = 0) = 0, \quad (\d P/\d r)_{r=0} = 0~~.
\ee
     The integrations are terminated when
     $P = 0$, which defines the surface $r = R$. A specified value
     of the
     central pressure $P_0=P(r = 0)$ determines the NS volume mass
     $M = m(r = R)$. Notice that the surface corrections to the
         NS volume mass
     were obtained in Ref.~\cite{MM24}.

     As shown in Ref.~\cite{RT87}, taking  approximately
     into account the step-function density,
     $\rho=\overline{\rho}$ for $r < R$ inside of the system, and $\rho=0$ for
     $r > R$ outside of it,
     one can solve
     TOV equations (\ref{toveq})
     explicitly analytically (see Refs.~\cite{RT87,MM24}).
         (For instance, $\overline{\rho}$ is the constant density $\rho_{00}$
   using units $G=c=1$, inside of the system
in the notations of Tolman).
     Our leading order solutions
         for the density $\rho$ are
         obtained in Ref.~\cite{MM24} by using the ESA
             within the leading leptodermic
         approximation; 
         see Eqs.~(31)$^\ast$ for more  general inter-particle interaction
         and (32)$^\ast$ for the specific vdW-Skyrme forces.
 They have a sharp transition from the saturation to zero
 value, which agrees with above mentioned approximation
     at zero order of the
 leptodermic expansion in small parameter
$a/R \ll 1$ (see Fig.~\ref{fig1}).
Taking also into account the first boundary condition
     in Eq.~(\ref{boundcondmp}), the second equation in
     Eq.~(\ref{toveq}) can be integrated
     analytically, 
     \be\l{mres}
     m(r)=\frac{4\pi\epsi_0}{3 c^2}r^3,
     \ee
     where $\epsi_0$ is a constant at $r<R$,
     \be\l{Eps0}
    \epsi_0=\epsi(\rho=
     \overline{\rho})~\Theta\left(\rho-\overline{\rho}\right)
     =\mathcal{A}(\overline{\rho})\Theta(R-r),
     \ee
     and $\overline{\rho}$ is the saturation approximation for the
     density $\rho$.
        We introduced here 
     the Heaviside step function, $\Theta(x)$, 
     $\Theta(x)=1$ for $x\geq0$ and $0$ for $x<0$.
      The function $\mathcal{A}(\rho)$ in Eq.~(\ref{Eps0})
is given by Eq.~(9)$^\ast$ of Ref.~\cite{MM24} 
for $\mathcal{A}(\rho)$ at the saturation density
$\rho=\overline{\rho}$; see Table \ref{table-1}.
     The density $\overline{\rho}$ and energy density $\epsi_0$ can be
     taken approximately as those found in Ref.~\cite{MM24}; see
     Eqs.~(3)$^\ast$, (9)$^\ast$, (10)$^\ast$, and (11)$^\ast$,
     including the macroscopic gravitational field. 
         Formally, one can consider the expression
        (9)$^\ast$ for $\epsi_0$ inside of the system
         [or Eq.~(16)$^\ast$] as the macroscopic
         EoS additional to the TOV equations
        (\ref{toveq}). 
Therefore,
 the total mass $M$ neglecting the surface effects and the influence
 of the Schwarzschild metric curvature,
 is given within this zero order ESA approximation by
     \be\l{Mtot}
     M=m(R)=\frac{4\pi\epsi_0}{3 c^2}R^3,
     \ee
     where $R$ is the effective NS radius ($r_1$ in the notations of
     Tolman's book, Ref.~\cite{RT87}).

    For the Schwarzschild metric in the interior region, one has
     \bea\l{Schwarz}
     &   {\rm d} s^2=-\frac{{\rm d} r^2}{1-r^2/R^2_{\rm S}}-
     r^2 {\rm d}\theta^2 -
     r^2 \sin^2\theta {\rm d} \phi^2\nonumber\\
     & + c^2\left[A_{\rm S} -
       B_{\rm S}\sqrt{1- r^2/R^2_{\rm S}}\right]^2 {\rm d} t^2~, 
     \eea
     where $R_{\rm S}$, $A_{\rm S}$, and $B_{\rm S}$ are the
         specific 
      integration
     constants, and $r<R_{\rm S}$.
      This metric, Eq.~(\ref{Schwarz}), is non-singular  at $r=0$,
      in contrast to the original  Schwarzschild metric \cite{RT87,LLv2},
     which is valid outside of the gravitating system.
      Therefore, it
      can be applied for the part of the space inside of
      this system
      (see Ref.~\cite{RT87}).
     A smooth joining of the external and
         internal Schwarzschild metrics through
     the effective radius, $r=R$, is assumed to be carried out. 
      The Schwarzschild radius $R_{\rm S}$ in Eq.~(\ref{Schwarz}) is
     expressed in terms of the
     energy density $\epsi$ inside of the system,
     $\epsi_0$,
     far away from
     the ES as  
     \be\l{RTOV}
     R_{\rm S}=\left(\frac{8\pi G\epsi_0}{3 c^4}\right)^{-1/2}.
     \ee
     Other constants, $A_{\rm S}$ and $B_{\rm S}$, as well as
     $R_{\rm S}$,
     are the following parameters of the 
     transformed Schwarzschild metric, Eq.~ (\ref{Schwarz})
      (see the derivations
     in Ref.~\cite{RT87}),
     \be\l{Tolconsschwm}
      A_{\rm S}=\frac32\sqrt{1-\frac{R^2}{R_{\rm S}^2}},\quad
      B_{\rm S}=\frac12~.
      \ee

  Substituting Eq.~(\ref{mres}) into the first TOV Eq.~(\ref{toveq}),
     one can see that the variables $P$ and $r$ are separated. Therefore,
     we
     analytically find  the solution
     for $r \leq R \leq R_{S}$: 
     \be\l{solTOVplus}
    P = \frac{\epsi_0}{3}
     \frac{3\zeta \sqrt{1-(r/R_{S})^2}-1}{1- \zeta\sqrt{1-(r/R_{S})^2}},
     \ee
            where
     \be\l{zeta}
     \zeta=\Big| \frac{P_0 + \epsi_0/3}{P_0 + \epsi_0}\Big|~,
     \ee
     and $P_0\equiv P(\overline{\rho})=P(r=0)$.
     Here, $R_{\rm S}$ is the radius
     parameter, Eq.~(\ref{RTOV}), in the 
         Schwarzschild metric, Eq.~(\ref{Schwarz}).
\begin{figure}
  \includegraphics[width=8.0cm]{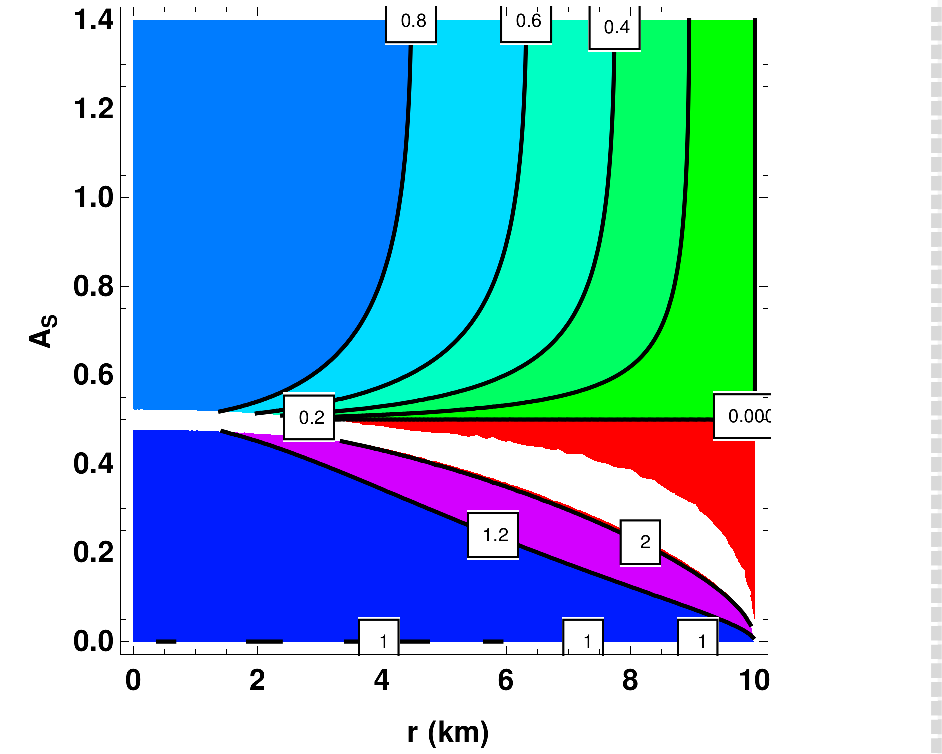}

  \vspace{-0.2cm}
  \caption{{\small
      Contour plots for the
      pressure
      $P(r)$, Eq.~(\ref{solTOVplus}),
      in units of 
      the central value, $P(r=0)$  
      as function of the radial
      coordinate $r$ and the dimensionless
      parameter $A_{\rm S}$ of the 
          Schwarzschild metric,
      Eq.~(\ref{Schwarz}). The numbers in squares show the values of
      this pressure.  White color presents regions where we have
          in-determination, infinity by infinity, with the finite limit 1
          at $r \rightarrow 0$. Red color shows negative values of the
          ratio 
          $P(r)/P(r=0)$ [a positive pressure
          $P(r)$]. 
          The value ``0.00'' displays the zero value in the
          horizontal and
      vertical coordinate lines on right of plots.
      The effective NS radius $R=10$ km is the same as in Fig.~ \ref{fig1}
      (from Ref.~\cite{MM24}). 
}}
\label{fig2}
\end{figure}
\begin{figure}
  \vskip1mm
  \includegraphics[width=8.7cm]{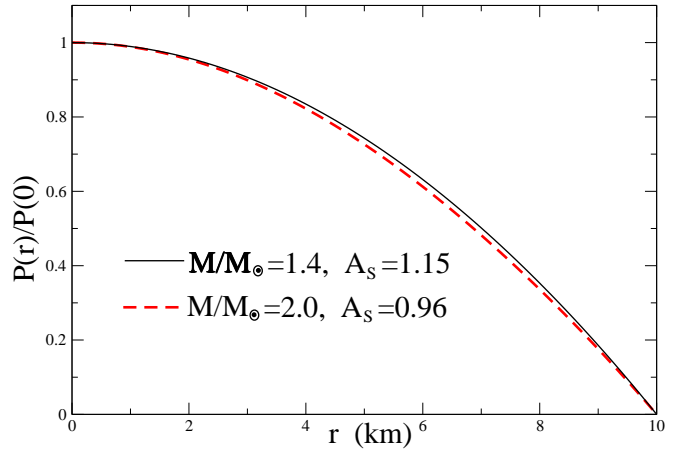}

  \vspace{0.2cm}
  \caption{{\small
   The pressure $P(r)$, 
      Eq.~(\ref{solTOVplus}),
      in units of 
     the central value, $P(r=0)$, 
      as function of the radial
      coordinate $r$ for the Schwarzschild
      metric, Eq.~(\ref{Schwarz}), at two 
          values of the parameter $A_{\rm S}$, $A_{\rm S}=1.15$
      (solid line)
          and $A_{\rm S} \approx 0.96$ (dashed line).
          The latter 
          corresponds to
          the two 
          values for the NS masses, $M=1.4 M_\odot$
      and $M=2.0 M_\odot$ ($R_{\rm S} \approx 13.0$ km and
      $15.6 $ km, respectively; see 
          Refs.~\cite{GR21,CC20,TR21,OL20} and text).
      The effective NS radius $R=10$ km is the same as in Figs.~
      \ref{fig1} and \ref{fig2}.      
}}
\label{fig3}
\end{figure}
     Notice that the solution, Eq.~(\ref{solTOVplus}),
     coincides, up to a constant, with
     that presented in Tolman's book at
     $\zeta=B_{\rm S}/A_{\rm S}=1/(2 A_{\rm S})$ 
     (see Refs.~\cite{RT87,MM24} and
     Eq.~(\ref{RTOV}) for $R_{\rm S}$).
      The cosmological constant also is
     assumed to be zero, and the units
     for the dimensionless pressure are used,
     $\epsi_0\propto R^{-2}_{\rm S}$, Eq.~(\ref{RTOV}).
    Thus, one obtains the relation of these parameters of the
      Schwarzschild metric, Eq.~(\ref{Schwarz}), to the initial condition
      $P=P_0$ at $r=0$ through Eq.~(\ref{zeta}) for $\zeta$. 
      Notice that according to 
      Eq.~(\ref{solTOVplus}) for the pressure
      $P$, one has $P=0$ at the boundary $r=R$
      because of
      Eq.~(\ref{Tolconsschwm}) for the constants $A_{\rm S}$ and $B_{\rm S}$.
      Finally,
      our result for the pressure, 
          Eq.~(\ref{solTOVplus}), coincides identically
     with that of
     Ref.~\cite{RT87}.

     Figure~\ref{fig2}
     shows the contour plots for the pressure $P(r)$
     [in units of 
      $P(r=0)$],
     as function of $r$ (km) and parameter $A_{\rm S}$,
     Eq.~(\ref{Tolconsschwm}). It is more convenient to use  the finite
     dimensionless  values of $A_{\rm S}$, $ 0 \leq A_{\rm S}\leq 1.5$,
     instead of dimensionless $R_{\rm S}/R$ for the fixed effective NS radius
     $R=10 $  km, $0\leq r \leq R $. For a very rough estimate
      neglecting the relatively small gravitational mass defect,
       $\epsi_0 \sim c^2\rho^{(0)}$,
        where the central mass density
        $\rho^{(0)} \approx M/(4\pi R^3/3) 
        \sim M_{\odot}/(4\pi R^3/3)$,
     ($M\approx M_V$)
     and $M_{\odot}=1.989*10^{30}$ kg is 
     the Sun mass,
     one finds significantly larger Schwarzschild radius parameter
     $R_{\rm S}$ than
     the effective radius $R$, $R_{\rm S} \approx 15.6$ km
      for $M=1.4 M_{\odot}$,
     as example.
     By definition, $R_{\rm S}$ is always larger
     than (or equal to) the effective NS radius $R$,
         $R\le R_{\rm S}$, while
     the gravitational
     radius $r_{\rm g}$ should be smaller than $R$,
     $r_{\rm g}=2GM/c^2 \leq R$ \cite{LLv2}. These conditions
     are fulfilled in our calculations
     for the values $R_{\rm S}\approx 15.6$ km,
     and $r_{\rm g}\approx 4.1$ km in the case of the effective radius $R=10$ km.
        With these
        evaluations, one has the constant $A_{\rm S} \approx 1.15$.
             The maximum value
        $A_{\rm S} \approx 1.5$  corresponds to
             zero pressure $P(r)$.
             
   Figure~\ref{fig3} shows the 
            two cuts of the contour plots
        (Fig.~\ref{fig2}) at the value,  $A_{\rm S}\approx 1.15$
        ($R_{\rm S}\approx 15.6$ km,
        solid black line),
        and the value,
        $A_{\rm S}=0.96$ ($R_{\rm S}=13.0$ km,  dashed red line). 
        The physical mass region,
        $ 
        M/M_{\odot}\approx 1.4-2.0$,
        in line 
            with the experimental
            data for NS masses and radii
            (see, Refs.~\cite{GR21,CC20,TR21,OL20}), is related to
            a small range
            in Fig.~\ref{fig3}. These curves have the same starting
            value 1 and final
            value 0 for the ratio of pressures $P(r)/P(0)$.
              For more realistic values of $R_{\rm S}$, for instance,
     taking into account the gravitational defect of mass \cite{LLv2} in
     the evaluations of the central energy density $\epsi_0$, one may have
     different values of $A_{\rm S}$ presented in Fig.~\ref{fig2}.
     Thus, as seen 
     from the solid and
             dashed lines of Fig.~\ref{fig3} and assumed in
     the derivations of the TOV equations \cite{OV39,RT87}, the
       pressure
         $P$ 
         turns, indeed, to zero
         at the boundary
     of the NS system, $r=R$.
  
\section{Modified TOV approach}
\l{mtov}

We now derive the leptodermic corrections to the TOV equations
due to the gradient
terms of the energy density $\epsi$, 
in line of Ref.\ \cite{MM24}.
The details of these derivations
are shown 
in Appendix \ref{appA}.

\subsection{General points}

The key quantity of our derivations is
    the energy density $\epsi\left(\rho\right)$ \cite{strmagden,BM15,MM24},
\be\l{enerden}
\epsi\left(\rho\right) =\mathcal{A}(\rho)
+\mathcal{C}\left(\nabla \rho\right)^2.
\ee
which determines the total system energy,
\be\l{energytot}
E=\int \d \mathcal{V}\; \epsi [\rho({\bf r})]~.
\ee
    The volume element, $ \d \mathcal{V}$, for a spherical system
    is determined by
    the Schwarzschild
    metric, $\d \mathcal{V}=\exp(\lambda/2) \d {\bf r}$, where
    $\d {\bf r}=r^2 \sin\theta~\d \theta \d \varphi $. 
        The integration over $\d \mathcal{V}$ is carried out for the whole
    infinite space.
   In Eq.~(\ref{enerden}), $\mathcal{A}(\rho)$ is the volume non-gradient part,
Eq.~(9)$^\ast$,
\be\l{vol0}
\mathcal{A}=- b^{}_V\rho + \eps(\rho)
+\frac{G}{4}\rho\Phi(\rho),
\ee
where $b^{}_V$ is the non-gravitational contribution into the
energy of separation of particle from the matter.
In Eq.~(\ref{vol0}),
\be\l{eps}
\eps(\rho)=\frac{K}{18 \overline{\rho}^2}~
\rho\left(\rho-\overline{\rho}\right)^2, 
\ee
where $K$ is the non-gravitational part of the
incompressibility modulus, e.g., due
to the Skyrme nuclear interaction (for nuclear
matter $K\sim 200$ MeV),
$\Phi(\rho)$ is the 
statistically
averaged (macroscopic) 
gravitational field; see Eq.~(8)$^\ast$ and discussions around it
in Ref.~\cite{MM24}.
For simplicity, we will consider the gradient term of the
inter-particle
interaction with a main constant
$\mathcal{C}$ 
    of  Eq.~(12)$^\ast$ for a smooth coefficient $\mathcal{B}(\rho)$
in front of the gradient-density square term.

We also use the condition for a minimum of the
energy per particle, $\mathcal{W}$, at a stable equilibrium,
\be\l{satcond}
\left(\frac{\d \mathcal{W}}{\d \rho}\right)_{\rho=\overline{\rho}} =0~,
\quad \mathcal{W}=\frac{\epsi}{\rho}~.
\ee
Therefore, when we can neglect external forces,
there is no linear terms in
expansion of $\mathcal{A}$,
Eq.~(\ref{vol0}), over powers of the difference
$\rho-\overline{\rho}$ near the saturation value $\overline{\rho}$.
Near the saturation value,
$\rho \rightarrow \overline{\rho}$, one has $\mathcal{W}=\mathcal{A}/\rho$.
Expanding the gravitational field $\Phi(\rho)$ near the saturation density
$\overline{\rho}$ up to second order, and taking into account
Eq.~(\ref{satcond}), one indeed
finds that there is no linear term, namely,
$\partial \Phi/\partial \rho=0$ at $\rho=\overline{\rho}$, i.e.,
$\Phi=\Phi_0+(1/2)\Phi_2(\rho-\overline{\rho})^2$ with some constants
$\Phi_0$ and $\Phi_2$. 
Combining different terms in Eq.~(\ref{vol0}), one writes
\be\l{vol}
\mathcal{A}=
-b^{(G)}_V\rho 
+\eps^{}_G(\rho),
\ee
where $b^{(G)}_V=b^{}_V - m \Phi_0$, $m$ is the test particle mass,
$\Phi_0=\Phi(\overline{\rho})$. 
The particle separation energy, $b^{(G)}_V$, and energy density component,
$\eps^{}_G(\rho)$
(see Eq.~(\ref{eps}) for $\eps(\rho)$) are
modified both
by the gravitational field
$\Phi(\rho)$, 
\be\l{epsKG}
\eps^{}_G(\rho)=\eps(\rho)+\frac{m}{2}
\Phi_2(\rho-\overline{\rho})^2
=\frac{K_G}{18 \overline{\rho}^2}~
\rho\left(\rho-\overline{\rho}\right)^2~. 
\ee
Here, $K_G$ is the total incompressibility modulus modified by the
gravitational field $\Phi(\rho)$ ($K_G>0$) as
\be\l{KG}
K_G=K+ 9 m \overline{\rho}^2\Phi_2~. 
\ee
and $\Phi_2$ 
is the second derivative of $\Phi(\rho)$,
$\Phi_2=\partial^2 \Phi/\partial \rho^2$, taken at the saturation
density value $\rho=\overline{\rho}$.

 For non-rotated one-component 
NS system
we
have to fix the particle number $N$,
\be\l{part}
N=\int \d \mathcal{V} \rho({\bf r})~.
\ee
Therefore, introducing the chemical potential $\mu$ as the Lagrange
multiplier, 
    one obtains from Eqs.~(\ref{energytot}) and (\ref{enerden})
the equation for the equilibrium,
\be\l{eq}
\hspace{-0.5cm}\frac{\delta \epsi}{\delta \rho}\equiv
\frac{\partial\mathcal{A}}{\partial\rho}
-2 \mathcal{C}\Delta \rho =\mu~.
\ee
Equation (\ref{part})
determines the chemical potential $\mu$ in terms of the
particle number $N$.  
For nuclear
liquid drop, one has two equations related to the two constraints for
the
fixed
neutron and proton numbers
 (Refs.~\cite{MS09,BM13,BM15}). 
They determine the two (neutron and
proton) chemical potentials.

The solution of the Lagrange equation (\ref{eq}) at leading order of the
leptodermic expansion over the parameter $a/R$, 
  $\rho(r)=\overline{\rho} y((r-R)/a)$, as function of
the radial coordinate $r$, 
is shown in Fig.~\ref{fig1} in the main case (i),
$\mathcal{B}=\mathcal{C}=const$,
for typical parameters,
the thickness of the
  NS diffused layer (crust) $a=1$ km and
  the effective radius $R=10$ km,
  in  units of the saturation value $\overline{\rho}$; see 
  Refs.~\cite{ST04,HPY07,MM24}.
  We compare the two dimensionless solutions for the 
  particle density $\rho(r)/\overline{\rho}$.
  One of them (solid line) is
  related to the vdW-Skyrme 
  interaction constant
  $\mathcal{C}$ ``1'' for a dense liquid drop;
  see Eq.~(32)$^\ast$ 
  for the
  universal dimensionless particle number
  density of a dense liquid drop,
  $y=\rho/\overline{\rho}$, as
  function of $x=(r-R)/a$. Another limit case corresponds to the
  symmetric Wilets solution (ii) \cite{wilets,MM24} 
   for zero constant $\mathcal{C}$
  and no spin-orbit term but with taking into account
  the only kinetic gradient term in a gas system; see
  Eq.~(12)$^\ast$
  for $\mathcal{B}$.
  These two curves are very different in the surface layer,
  especially outside of the system far away from the NS, and almost the same
  inside of the NS.
   As seen from Fig.~\ref{fig1} and expected, the
  range of the exponential decrease of $\rho(r)$ becomes much smaller
  for the vdW-Skyrme liquid-drop case (i) than for the Wilets gas
  case (ii). But
  the surface effects on the local density $\rho(r)$ in both cases
  are still very
  important to
  get a continuous analytical behavior through the ES.
  The smallness factor, proportional to $a/R$, with
  respect to the volume contribution, appears
    because of the integration over the normal-to-ES
  (radial in this example) variable $x=(r-R)/a$ in
  this ES region.  Thus, 
      for the calculations of the surface corrections to the
NS total energy $E$,
Eq.~(\ref{energytot}), and particle number $N$, Eq.~(\ref{part}),
it would seem possible to assume 
that they
are not (very) important 
for
so small relative diffuseness thickness $a/R=0.1$.
However, as shown in Ref.~\cite{MM24},
the influence of the NS surface becomes really important
for the local
  quantities as the particle density $\rho(r)$, 
      Fig.~\ref{fig1}, the EoS,
      and therefore, for the pressure $P(r)$ 
          calculating through
          Eqs.~(\ref{enerden}) and (\ref{satcond}),
\be\l{EoS}
P=\rho^2 \frac{\delta \mathcal{W}}{\delta \rho}~.
\ee
Moreover, the gravitational mass defect due to the radial curvature
of the nonlinear Schwarzschild metric in calculations of the NS mass
leads to its dramatic non-monotonic behavior of $M=M(R)$; see Ref.~\cite{MM24}.
\begin{figure*}
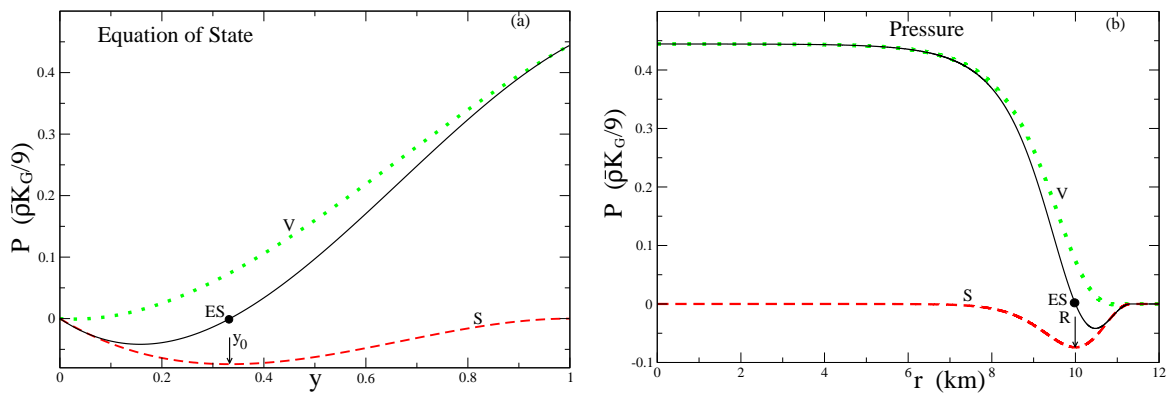

  \vskip1mm
  \includegraphics[width=7.5cm]{Fig4a-PVStoty.eps}
  ~
    \includegraphics[width=7.5cm]{Fig4b-PVSr_a1R10.eps}

  \vskip5mm\caption{{\small
      Pressure $P$,
      Eq.~(60)$^\ast$, 
      in units of 
      $ \overline{\rho} K_G/9$ 
      through the NS diffuse surface as function of
      the density variable $y$, (a), for the universal Equation of State,
      and the radial
      coordinate $r$, (b), 
      through the particle number
          density, $\rho=\overline{\rho}y((r-R)/a)$. The density
          $y(x)$ is given, e.g., by Eq.~(32)$^\ast$
          (solid lines). Dashed and dotted lines show the surface (S)
              and volume (V) components, $P_{\rm S}$ [Eq.~(62)$^\ast$]
                  and $P_{\rm V}$
                  [Eq.~(61)$^\ast$], respectively.
      The crust thickness $a=1.0$ km 
      and the effective radius parameter $R=10$ km
      are the same as in Figs.~\ref{fig1} and \ref{fig2}.
      The full dots and arrows present the ES for $y=y_0$, (a),
      and $r=R$, (b).    
}}
\label{fig4}
\end{figure*}

Figure~\ref{fig4} shows the
 pressure $P$ as function of the dimensionless density $y$ 
     [panel (a)]
 and radial variable $r$ [panel (b)]
 through the density, $y((r-R)/a)$,
at leading order over
the parameter $a/R$; see Eq.~(60)$^\ast$ 
for the full pressure ($P$),
Eq.~(61)$^\ast$ 
for the volume ($P_{\rm V}$), and Eq.~(62)$^\ast$ 
for the surface ($P_{\rm S}$)
components of the pressure
(Table~\ref{table-1} and Ref.~\cite{MM24}).
As seen from Fig.~\ref{fig4} (a),
all pressures, $P$, $P_{\rm V}$ and $P_{\rm S}$, take zero value at $y=0$.
The surface pressure, $P_{\rm S}$, takes zero value also asymptotically
in the limit $y \rightarrow 1$, in contrast to the volume part, $P_{\rm V}$,
as well as the total pressure $P$.
The full dot shows the ES at $y=y^{}_0=1/3$ for
the (i) case. As expected,
the volume pressure  $P_{\rm V}$ is monotonically
increasing function of the density $y$. In Fig.~\ref{fig4} (b), the pressure
$P(\rho(r))$ (solid line)
has a typical leptodermic behavior with a relatively
short thickness of the order of $a$,  where the pressure decreases sharply
from the almost constant asymptote inside of the system at
$r \simg R-a$  to zero. Dashed red and dotted lines show
the surface ($P_{\rm S}$) and volume ($P_{\rm V}$) components, respectively.
Sum of these components is the total pressure $P$ with zero 
    value at the ES
because of the equilibrium; see
the full point (ES) and arrow which present the effective surface and
radius $R$, respectively.
      Taking the positive direction from the inner
      $(r<R)$ to outer $(r>R)$ region, one has
      a positive volume pressure $P_{\rm V}$ for any dense  liquid drop.
      The surface capillary pressure $ P_{\rm S} $ is negative because
      there is a dense matter inside $(\rho \approx \overline{\rho} >0)$
      and almost no stellar matter  outside
      of the NS [$\rho$ tends exponentially  to zero 
      in the outer region $r>R$; see, e.g., Ref.~\cite{MM24}],
      i.e. $d\rho/dr <0$. Therefore, there is an opposite capillary force
      directed to the interrior of the system \cite{RW82,LLv6}.
      When the resulting pressure
      $P>0$, the system is expanded, and for $ P<0$ the liquid-drop system 
      is compressed, in contrast to
      gas systems. All these macroscopic properties are directly proved 
      analytically from the
      definition for the EoS, Eq.~(\ref{EoS}) for the pressure, Eq.~(\ref{satcond})
      for the energy density per particle $W$ and explicit expressions
       (61)$^\ast$ for the volume $P_V$
 and (62)$^\ast$ for the surface $P_S$ pressures; see Table \ref{table-1}.

\subsection{Surface corrections to the TOV equations}

For the energy density $\epsi$, Eq.~(\ref{enerden}), one can write
\be\l{enerden1}
\epsi=\epsi_0+\epsi_1~,
\ee
where $\epsi_0$ is given by Eq.~(\ref{Eps0}); see 
Eqs.~(3)$^\ast$ and (12)$^\ast$ (Table~\ref{table-1}). The second component
can be considered as a small leptodermic
correction,
\be\l{e0e1}
\epsi_1= \mathcal{C}\left(\nabla\rho\right)^2.
\ee
Indeed, $\epsi_1$
enters the solutions of Eq.~(\ref{lamdanueq}) in terms of the integral
over the radial coordinate $r$.
Our derivations at leading order over leptodermic parameter
$a/R$ are similar to those for the equation for the density $\rho$
and its solutions in
Ref.~\cite{MM24}.

Starting from the same equations (\ref{lamdanueq}), as presented in the
previous section (see Ref.~\cite{RT87}),
we split all their solutions into two components, the leading component
and leptodermic corrections,
\be\l{01}
P=P_0+P_1,\quad \lambda=\lambda^{}_0+\lambda^{}_1, \quad \nu=\nu^{}_0+\nu^{}_1~,
\ee
where $P_0$ is the zero order solution shown in
Eq.~(\ref{solTOVplus}).
It is related to
the constant saturation value for the
density $\overline{\rho}$ inside of the system and zero
outside, and
corresponding energy
density constant
$\epsi_0$; see Eq.~(\ref{Eps0}). These quantities correspond to
the zero order
Schwarzschild metric parameter, $\lambda^{}_0(r)$, which is the
function of $r$ (see Eq.~(\ref{ds2schw}) and Ref.~\cite{RT87}),
\be\l{lam0}
e^{-\lambda^{}_0}=1-r^2/R^2_{\rm S}~, \quad r < R_{\rm S}~.
\ee
The component $\nu^{}_0(r)$ 
for the Schwarzschild metric
(\ref{ds2schw}) can be found similarly as done in
Ref.~\cite{RT87}. The leptodermic corrections, $P_1$ and $\nu^{}_1$,
Eq.~(\ref{01}),
can be expressed in terms of the $\lambda^{}_1$.

As shown in Appendix \ref{appA}, 
    using approximately the property of the energy density component
    $\epsi_1\propto (\partial \rho/\partial r)^2$, which is
    concentrated
    near the ES, for the correction $\lambda^{}_1$, at first order
in the leptodermic expansion, one finally obtains
from the second equation of Eq.~(\ref{lamdanueq}),
\be\l{sollam1final}
\lambda^{}_1(r)=-\frac{8\pi R_{\rm S}}{1-R^2/R_{\rm S}^2}~\mathcal{D}_1(r)~,
\ee
where
\be\l{e1intdef}
\mathcal{D}_1 \equiv \int {\rm d} r\epsi_1(r) 
\approx \frac{a}{R}
\frac{R K_G}{9\overline{\rho}}\int \d \xi
\left(\frac{\partial \rho}{\partial \xi}\right)^2~.
\ee
 The integrations over $r$ and $\xi=r-R$ are indefinite, 
as well as in all equations below,
    in particular in Appendix \ref{appA}.

Substituting the asymptote of the solutions, $\rho \sim \exp(-\xi/a)$,
for larger distances  $\xi/a \gg 1$ (see Ref.~\cite{MM24}),
one has an exponential disappearance
of the integral $\mathcal{D}_1$, Eq.~(\ref{e1intdef}), 
\be\l{e1intout}
\mathcal{D}_1\propto \exp\left(-2\xi/a\right)~.
\ee
For the inner region,
one finds also the exponentially decreasing asymptotes
of $\mathcal{D}_1$, Eq.~(\ref{e1intdef}),
\be\l{e1intin}
\mathcal{D}_1\approx -\frac{a}{R}~ \frac{\overline{\rho} R K_G}{9}
 \mathcal{Y}(y).
\ee
The function $\mathcal{Y}$ is given
by
\bea\l{Y}
&\mathcal{Y}(y)=\int {\rm d}y \sqrt{y}(1-y)\nonumber\\
&=\frac{2}{3}y^{3/2}-\frac{2}{5}y^{5/2} + C_y~,
\eea
where
$y=\rho/\overline{\rho}$  (see Ref.~\cite{MM24}),  $C_y$ is an
integration constant,   and $\d y=\d r(\d y/\d x)/a$ with
$y=y(x)$ of Eq.~(32)$^\ast$
in the integral,
Eq.~(\ref{Y}).
 To remove singularity at the origin, we
 shall assign the value $-4/15$ for the 
 constant $C_y$.
    Notice that  in the derivations for $\lambda^{}_0$, Eq.(\ref{lam0})
    [Eq.~(96.7) in Tolman's
      book \cite{RT87}] similar corrections were neglected
    by the same arguments.
    The solution $\lambda^{}_1(r)$,
    Eq.~(\ref{sollam1final}), is zero outside of the system because
    of zero $\epsi_0$, Eq.~(\ref{Eps0}), and $P_0$, Eq.~(\ref{solTOVplus}).
    Therefore, $\lambda^{}_1(r)$
    is concentrating 
    near the NS surface, as it should be true (see also Appendix \ref{appA}).

Substituting the derivative ${\rm d} \nu/{\rm d} r$ from the first equation of
Eq.~(\ref{lamdanueq}) to its last equation, 
for a given energy density $\epsi$, one obtains the modified TOV equation
for the pressure $P$,
\be\l{MTOV}
\frac{{\rm d} P}{{\rm d} r}=-\frac{\epsi+P}{2r}~
\left[\left(8\pi P r^2+1\right) e^\lambda-1\right],
\ee
where $\lambda = \lambda^{}_0+\lambda^{}_1$; see
Eqs.~(\ref{lam0}) and
(\ref{sollam1final}).  Notice that for shortness, from this
equation to the final result, Eq.~(\ref{solP1fin1}), and in Appendix \ref{appA} we use
the generally accepted units $c=G=1$, in particular as in Ref.~\cite{RT87}.
 Equation (\ref{MTOV}) is valid
approximately up to high, e.g. second order in the
leptodermic expansion over $a/R$.
Let us use now Eqs.~(\ref{enerden1}) for $\epsi$ and
(\ref{01}) for $P$. Using also Eq.~(\ref{eq2lam0}) for
$P_0$ at zero order of the leptodermic parameter
$a/R$, one can derive the
linearized equation for $P_1$ 
    beyond the analytical solutions (\ref{lam0}) and (\ref{solTOVplus}) of
the standard TOV equations
for $\lambda^{}_0$ and 
$P_0$,
respectively.

For the leptodermic correction $P_1$ to the pressure $P$,
one has the same standard form as Eq.~(\ref{eq2lam1})
for $\lambda^{}_1$,
\be\l{eqP1}
\frac{{\rm d} P_1}{{\rm d} r} + \beta P_1=\beta^{}_1~.
\ee
Here,
$\beta(r)$ and $\beta^{}_1(r)$ are new
functions of the radial variable $r$,
\be\l{qRs}
\beta=\frac{8\pi r^2 \left(\epsi_0+P_0\right)+
8 \pi P_0 r^2 +\tilde{r}^2}{2r \left(1-\tilde{r}^2\right)}~,
\ee
\be\l{sRs}
\!\beta^{}_1\!=\!-\frac{8 \pi P_0 r^2 \!+\! \tilde{r}^2}{2r \left(1-\tilde{r}^2\right)}
\epsi_1 
\!+\!\frac{\left(\epsi_0\!+\!P_0\right)\left(8\pi P_0 r^2 +1\right)}{1-\tilde{r}^2}
\lambda^{}_1~,
\ee
where $\tilde{r}=r/R_{\rm S}$, as in Appendix \ref{appA}.
In these equations, 
$P_0(r)$ is the analytical solution
of the TOV equations at zero
order of the leptodermic parameter for the
pressure, Eq.~(\ref{solTOVplus}); see Refs.~\cite{RT87,MM24}.
Using the same method of the arbitrary constant variations,
one obtains the solution in terms of the
 indefinite integrals quadratures (see Appendix \ref{appA}),
\be\l{solP1}
P_1=-C_1^{(P)}(r) \exp\left(-\int {\rm d}r \beta(r)\right),
\ee
where
\be\l{solCp}
C_1^{(P)}(r)=\int {\rm d}r \beta^{}_1(r)
\exp\left[\int {\rm d}r \beta(r)\right].
\ee

\subsection{Leading solutions for the pressure}
\l{sol}

Substituting the function $\beta(r)$, Eq.~(\ref{qRs}),
into the indefinite integrals for the pressure correction $P_1$,
Eqs.~(\ref{solP1}) and
(\ref{solCp}),
one finds 
\be\l{qintRs}
\exp\left[{\int {\rm d} r \beta(r)}\right]
=\frac{\left(1-\zeta\sqrt{1-\tilde{r}^2}\right)^2}{\sqrt{
      \tilde{r}\left(1-\tilde{r}^2\right)}}~,
\ee
where $\zeta= 1/(3\sqrt{1-\tilde{R}^2})$ 
    with $\tilde{R}=R/R_{\rm S}$.
Using then Eq.~(\ref{sRs}) for $\beta^{}_1(r)$ in
Eq.~(\ref{solCp}) for $C_1^{(P)}$, one obtains the sum of the two terms,
\bea\l{solCp1}
&C_1^{(P)}(r)=-\int {\rm d}r
\left[\frac{8 \pi P_0 r^2 + \tilde{r}^2}{2r \left(1-\tilde{r}^2\right)}
\epsi_1 \right.
\nonumber\\
&+\left.\frac{\left(\epsi_0+P_0\right)
  \left(8\pi P_0 r^2 +1\right)}{1-\tilde{r}^2}
\lambda^{}_1\right]~\exp\left[\int {\rm d}r \beta(r)\right].
\eea
The integrand of one of  such integrals is proportional to $\epsi_1$,
Eq.~(\ref{e0e1}), while the second one is proportional to $\lambda^{}_1$,
Eq.~(\ref{sollam1final}).
The first order corrections,
$\epsi_1$ and $\lambda^{}_1$, are determined
by Eqs.~(\ref{e0e1}) and
(\ref{sollam1final}), respectively
[see also the approximate Eq.~(\ref{e1})].
Both these functions exponentially
decrease for $r$ 
outside of a relatively small range $a$ of the NS crust
at small
leptodermic parameter, $a/R$.  
Therefore, the first integral in Eq.~(\ref{solCp1}) 
can be taken approximately at leading order, $a/R$, in the
leptodermic expansion as for the derivation of $\lambda^{}_1$.
Notice that 
the second integral in Eq.~(\ref{solCp1}), with the integrand
$\propto \lambda^{}_1(r)$, over radial coordinate $r$ can be
neglected as the second order
    of $(a/R)^2$ in the leptodermic expansion. Therefore,
it does not contribute
at first order over $a/R$. Indeed, this 
is reduced approximately
to $\int {\rm d} r \mathcal{Y}(y)$,
\bea\l{intlam}
&\int {\rm d} r \lambda^{}_1(r)\approx -
\frac{a}{R} \frac{K_G\overline{\rho}}{9}
\frac{4\pi R^2}{\sqrt{1-R^2/R_{\rm S}^2}}\nonumber\\
&\times\int {\rm d} r \mathcal{Y}(y).
\eea
    Notice that by the same reason, the $\lambda^{}_1$ surface correction to the
mass defect owing to the curved space can be neglected in calculations of
the NS mass $M$ (Ref.~\cite{MM24}) at the first order over $a/R$.
Transferring the last integral over $r$ to the variable $\rho$ and then to
the dimensionless $y$, as above,
${\rm d} r= a {\rm d} y/({\rm d}y/{\rm d}x)$,  and using
Eqs.~(27)$^\ast$, and (19)$^\ast$
for the inverse derivative of $y$ over $x$,
one finds
\be\l{intY}
\int {\rm d} r \mathcal{Y}(y)\approx -\frac{a}{R} R
\int \frac{{\rm d}y}{\sqrt{y}(1-y)} \mathcal{Y}(y).
\ee
The  indefinite integral on the right hand side
is finite because its integrand has no singularity in
the limit $y\rightarrow 1$ by the
 L'Hopital
theorem, $\mathcal{Y}/(1-y) \rightarrow $ 
$\mathcal{Y}^{\prime}(y)/(-1) \rightarrow (1-y) \rightarrow 0~$.
There is also no singularity at small $y$,
$\int{\rm d} y/ \sqrt{y} =2\sqrt{y}$.

Taking all smooth quantities in front of $\epsi_1$ off the first
integral in Eq.~(\ref{solCp1}) over the radial variable $r$
at $r=R$,
one approximately finds
\be\l{solCpRs1}
C_{1}^{(P)}(r)\approx-\frac{\tilde{R}^2}{2R \left(1-\tilde{R}^2\right)}~
\mathcal{D}_1~,
\ee
where $\mathcal{D}_1$ is given by Eq.~(\ref{e1intdef}) with the same
asymptotic properties (\ref{e1intout}) and (\ref{e1intin})
outside of the narrow NS crust, and again, $\tilde{R}=R/R_{\rm S}$.
With Eq.~(\ref{qintRs}) for the integral exponent taken at $r=R$
and the coefficient $C^{(P)}_{1}$, Eq.~(\ref{solCpRs1}), one
arrives at the
final analytical approximation for the
surface pressure correction $P_1$,
\be\l{solP1fin}
P_1=-\frac{3 R^{5/2}}{4 R_{\rm S}^{5/2} \sqrt{1-R^2/R^2_{\rm S}}}~
\frac{\mathcal{D}_1}{R}~;
\ee
see Eq.~(\ref{solP1}).
 This final result (and below)
    is presented in terms of the coefficient $\mathcal{D}_1/R$ which
     has the clear pressure
    dimension [see Eq.~(\ref{e1intdef})]. Another factor is dimensionless
    as that expressed through the ratio $R/R_{\rm S}$ of the effective radius $R$ to the
    Schwarzschild radius $R_{\rm S}$, Eq.~(\ref{RTOV}), 
  due to the Schwarzschild
    gravitation. Therefore, taking Eq.~(\ref{RTOV}) into account,
    one can easy come back from the $c=G=1$ to the normal units.
The Schwartzschild radius $R_{\rm S}$,
Eq.~(\ref{RTOV}),
is related to the initial value
of the energy density, $\epsi_0$, and $\mathcal{D}_1$,
 Eq.~(\ref{e1intdef}), has the internal
(\ref{e1intin}) and external (\ref{e1intout}) asymptotes.
Therefore, these leptodermic corrections for $C_1^{(P)}$, and hence, for
$P_1$, (as $\propto \mathcal{D}_1$), are
exponentially decreasing functions of radial coordinate  $r$,
$r \rightarrow \infty$ outside [see Eq.~(\ref{e1intout})]
and to zero inside of the system [Eqs.~(\ref{e1intin}) and (\ref{Y})].
In particular, inside of the system, Eq.~(\ref{e1intin}) for
$\mathcal{D}_1$, one arrives at
\be\l{solP1fin1}
P_1=\frac{a}{R}~\frac{\overline{\rho}K_G}{12}~
\frac{R^{5/2}}{R^{5/2}_{\rm S}\sqrt{1-R^2/R^2_{\rm S}}}~\mathcal{Y}(y)~,
\ee
 with the explicitly presented pressure dimension.
Thus, as expected, the leading leptodermic correction $P_1$ to the zero order
pressure $P_0$   is peaked at the NS effective-surface range
due to the 
function $\mathcal{D}_1$ [Eqs.~(\ref{e1intdef})-(\ref{e1intout})].
Therefore,
one finds the equation of state in the form:
\be\l{EoS01}
P\equiv P(\rho) = P^{}_0(y)+P^{}_1(y),
\ee
where $y=\rho/\overline{\rho}$, $P^{}_0$ and $P^{}_1$ are given approximately,
at the first order leptodermic approach, by Eqs.~(\ref{solTOVplus})
and (\ref{solP1fin}), respectively.
Substituting the solutions $y=y(r)$ [$x=x(y)$, Eq.~(31)$^{\ast}$
with explicitly given Eq.~(32)$^{\ast}$]
into
Eq.~(\ref{EoS01}) we obtain $r$ coordinate dependence
of the pressure $P(\rho(r))$ at first order in $a/R$.
\begin{figure}
  \vskip1mm
  \includegraphics[width=8.0cm]{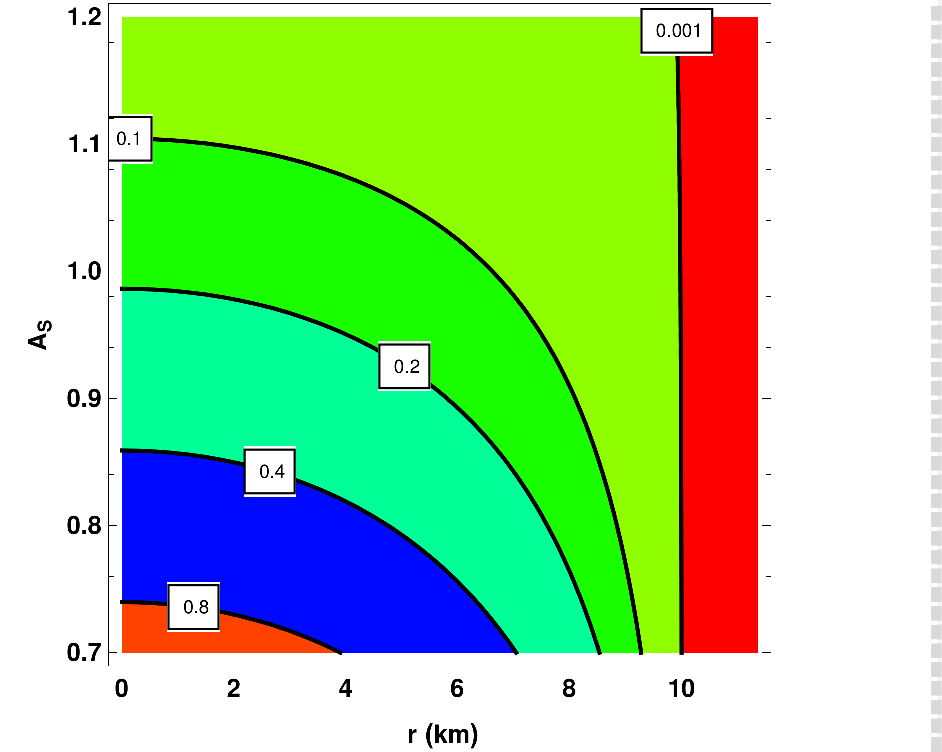}

  \vskip5mm\caption{{\small
      Contour plots for the dimensionless
      pressure
      $p^{}_0(r,A_{\rm S})$, Eq.~(\ref{p0cal}),
      as function of the radial
      coordinate $r$ and the dimensionless
      parameter $A_{\rm S}$ of the 
          Schwarzschild metric,
          Eq.~(\ref{Schwarz}). 
          Other notations
          and parameters are the same as in Fig.~\ref{fig2}.
          }}
\label{fig5}
\end{figure}
\begin{figure}
  \vskip1mm
    \includegraphics[width=8.0cm]{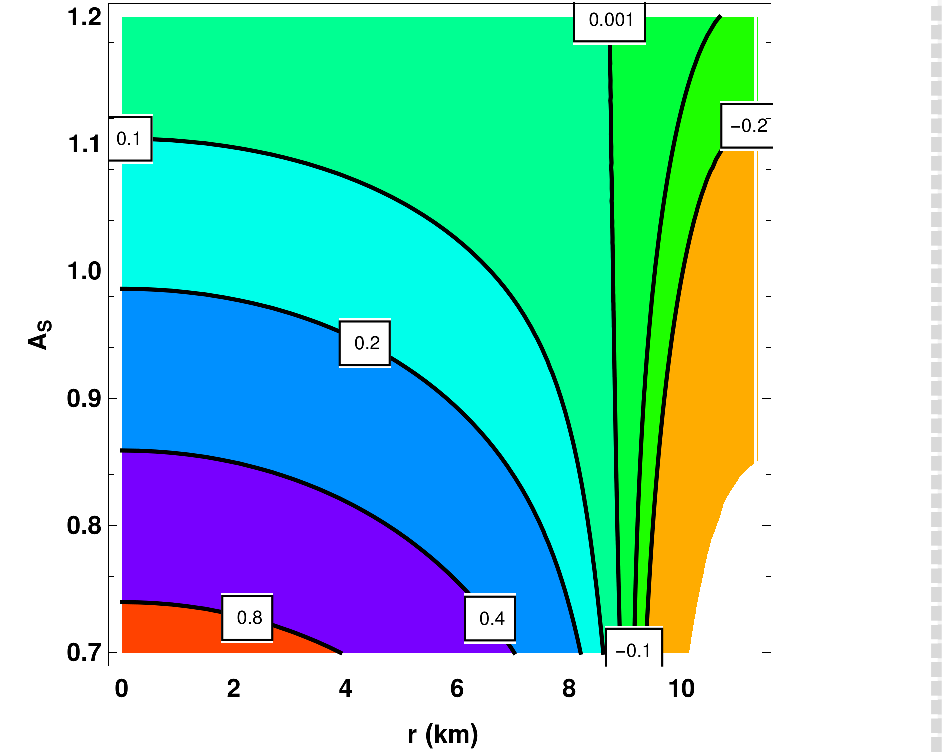} 

  \vskip5mm\caption{{\small
      The same as in Fig.~\ref{fig5} but for
      the total pressure $p(r,A_{\rm S})$,  Eq.~(\ref{p01cal}).
             White color presents regions where we have
          zero values.
          The effective NS radius $R=10$ km
          is the same as in
          Figs.~\ref{fig1}-\ref{fig4}. The leptodermic parameter
          has a typical value $a/R=0.1$. For example, the 
              dimensionless
          incompressibility parameter
          $\kappa=10$ [see Eq.~(\ref{kappa})]
               for large gravitational forces
      ($\kappa \gg 1$). 
}}
\label{fig6}
\end{figure}
\begin{figure}
  \vskip1mm
  \includegraphics[width=8.0cm]{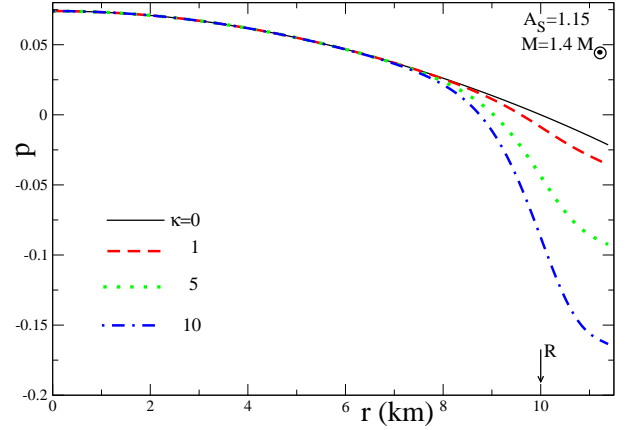}

  \vskip5mm\caption{{\small
 The dimensionless pressure 
     $p=P(8\pi GR^2/c^4)$, Eq.~(\ref{p01cal}),
      as function of the radial
      coordinate $r$ for the Schwarzschild
      metric in the form of Eq.~(\ref{Schwarz}) at   
      the same value of the parameter $A_{\rm S}$, $A_{\rm S}\approx 1.15$,
      and different values of the incompressibility
      parameter $\kappa=0$ [black solid for
      the zero order pressure
      $p^{}_0$, Eq.~(\ref{p0cal})], $1$
      [red dashed line, $5$ (green dotted line),
      and
      $10$ (blue dash-dotted line), Eq.~(\ref{p01cal})]. The
      NS mass is the same as
      for the solid line  in Fig.~\ref{fig3}, $M=1.4 M_{\odot}$.
      Other parameters are
      the same as in Figs.~\ref{fig5} and \ref{fig6}.               
}}
\label{fig7}
\end{figure}
\begin{figure}
  \vskip1mm
  \includegraphics[width=7.0cm]{Fig8_as096_k0-10_R10_aR01.eps}

  \vskip1mm\caption{{\small
      The same as in Fig.~\ref{fig7} but for the 
       NS mass 
        presented by the dashed line  in Fig.~\ref{fig3},
      $M=2.0 M_{\odot}$ ($A_{\rm S}=0.96$).         
}}
\label{fig8}
\end{figure}

It is also
convenient to introduce the following dimensionless pressures
as functions of $r$ and $A_{\rm S}$ variables, which are explicitly related
to the constants $c$ and $G$,
\be\l{p0cal}
\!p^{}_0\!\equiv\! \frac{8\pi P_0 GR^2}{c^4}\!=\!
\frac{R^2}{R_{\rm S}^2}
 \frac{
   3\sqrt{
     1\!-\!r^2/R_{\rm S}^2}/2\!-\!A_{\rm S}}{A_{\rm S}\!-\!\sqrt{
     1\!-\!r^2/R_{\rm S}^2}/2}~.
\ee
The explicit relation of the ratio $R/R_{\rm S}$ to $A_{\rm S}$ is given by
$R/R_{\rm S}=\sqrt{1-4A_{\rm S}^2/9}$; see Eqs.~(\ref{Tolconsschwm}) and
(\ref{RTOV}).
Then, for the correction $P_1$, one similarly writes
\be\l{p1cal}
\!p^{}_1\!\equiv\! \frac{8\pi P_1 GR^2}{c^4}
\!=\!\frac{3 \kappa R^{9/2}}{R^{9/2}_{\rm S}\sqrt{
    1\!-\!R^2/R^2_{\rm S}}}\mathcal{Y}(y)
\frac{a}{R}~,
\ee
where
  \be\l{kappa}
  \kappa\equiv\frac{\overline{\rho}K_G}{12\epsi_0}=
  -\frac{K + 9 m \overline{\rho}^2 \Phi_2}{12(b^{}_V - m \Phi_0)}~,
  \ee
  $K$ is a non-gravitating (e.g., nuclear)
  incompressibility;
    see also Eqs.~(\ref{Eps0}) for $\epsi_0$ and
  (\ref{vol}) for a smooth part of the
  energy density $\mathcal{A}(\rho)$ in terms of 
  the gravitational field derivatives of $\Phi$.
For the total dimensionless pressure $p$,
at first order in the leptodermic parameter $a/R$, one finally obtains
\be\l{p01cal}
p\approx p^{}_0+p^{}_1~,
\ee
where $p^{}_0$ and $p^{}_1$ are given by Eqs.~(\ref{p0cal})
and (\ref{p1cal}).

\section{Discussions of the results}
\l{disc}

Figures~\ref{fig5} and \ref{fig6}  show the dimensionless pressure
$p^{}_0$, Eq.~(\ref{p0cal}), at zero order
and the total pressure $p$, Eq.~(\ref{p01cal}), at the first order
over a leptodermic parameter
$a/R=0.1$
  as function of the radial coordinate $r$ and Schwarzschild metric
  constant $A_{\rm S}$, relatively.
  The first order contribution $p^{}_1$ of the pressure
  $p$ is proportional
  to the dimensional incompressibility parameter $\kappa$, Eq.~(\ref{kappa}).
  As seen from the comparison of Figs.~\ref{fig6}
  ($\kappa=10$)
  and \ref{fig5} ($\kappa=0$),
      the effect of
      the gravitational
      field on the pressure is significant 
          near the ES ($r\simg R$)
      for this change
       because of the ES correction $p^{}_1$,
          Eq.~(\ref{p1cal}) at a large
      dimensionless incompressibility $\kappa$, Eq.~(\ref{kappa}).

  Figures~\ref{fig7} and \ref{fig8} show the cuts of
  the contour plot of Fig.~\ref{fig6} at $A_{\rm S}\approx 1.15$
  corresponding approximately to the typical NS mass $M=1.4M_{\odot}$
  and $A_{\rm S}\approx 0.96$ for $M=2.0M_{\odot}$, respectively,
  as function of the radial coordinate $r$ (in km). These NS masses
  $M$ and approximately the radii by the order of the magnitude
  were found in the recent experiments presented in
  Refs.~\cite{GR21,CC20,TR21}; see also the comparison
  with the theoretical results in Ref.~\cite{OL20}. We specify
  different values of the
  incompressibility parameter $\kappa$ from a weak (almost nuclear
  $\kappa=1$) to a strong
  value ($\kappa=10$) of the gravitational
   field. As mentioned above, the value $\kappa=0$ is
  related to the pressure $p^{}_0$ for the zero order
  leptodermic parameter $a/R$; see  Fig.~\ref{fig5}.
  For example, one can consider stronger gravitational
  field up to the value $\kappa=10$
  in Figs.~\ref{fig7} and \ref{fig8}. For nuclear physics, one has
  $\kappa\equiv- K/(12b^{}_{\rm V})=1.25$
  for $K=240 $ MeV, and $b^{}_{\rm V}=-16$ MeV (or $K=225$ MeV and
  $b^{}_V=-15$ MeV
  from Ref.~\cite{AB14}).
     \begin{figure*}
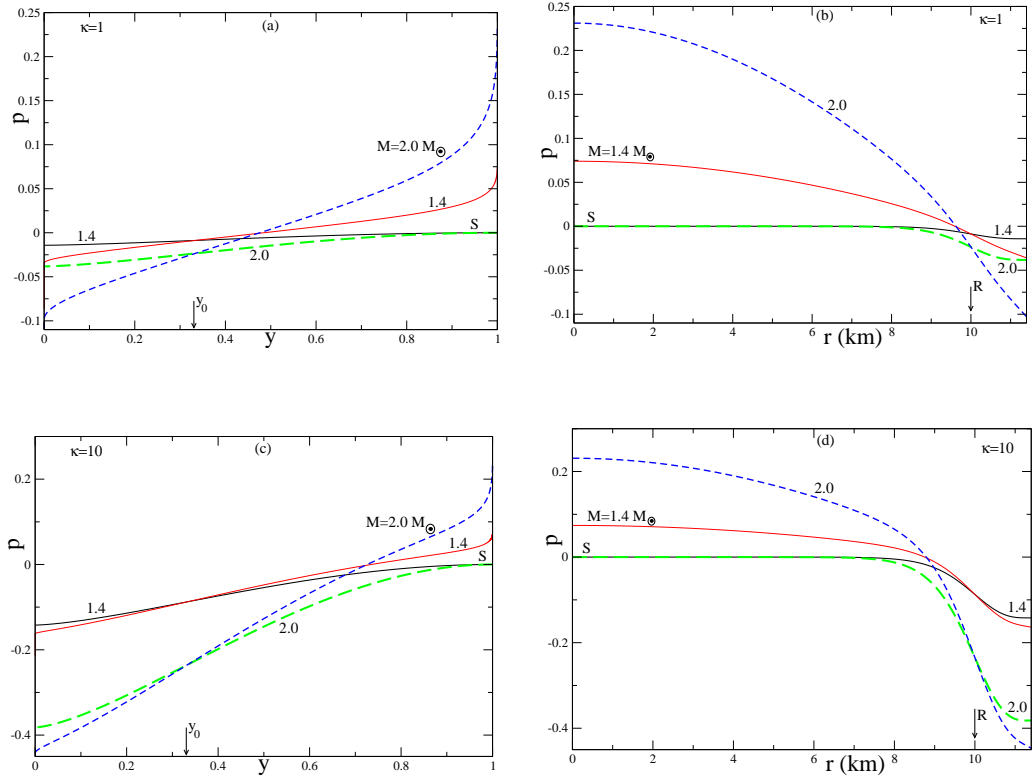

       \vskip1mm

 \includegraphics[width=6.5cm]{Fig9a_EoS_k1_R10_aR01.eps}
       ~~
        \includegraphics[width=6.5cm]{Fig9b_NSPSwar_k1_R10_aR01.eps}
       
       \vspace{1.0cm}
          \includegraphics[width=6.5cm]{Fig9c_EoS_k10_R10_aR01.eps}    
       ~~~
       \includegraphics[width=6.5cm]{Fig9d_NSPSwar_k10_R10_aR01.eps}

  \vspace{-1.0cm}
  \vskip5mm\caption{{\small
      The dimensionless pressures, $p$, Eq.~(\ref{p01cal}),
      (thick solid black line for the surface component $p^{}_1$ and
      thin red one for the total volume and surface pressure $p$)
      as functions of the dimensionless particle number density, $y$,
      (density $\rho$ in  units of $\overline{\rho}$)
      for the solution (32)$^*$ of leading order
      over the leptodermic parameter $a/R$, (a,c), and over the radial
      coordinate $r$, (b,d), for different dimensionless
      incompressibilites $\kappa=1$, (a,b), and $10$, (c,d),
      [see Eq.~(\ref{kappa})],
      for the NS mass $1.4M_\odot$.
      Dashed (rare 
          green  $p^{}_1$ and frequent blue $p$ ) lines
      show these pressures for the NS mass $2.0 M_\odot$.
       The arrows show the ES value
      $y^{}_0=1/3$ in (a,c) and the effective radius $R$ in (b,d).
      }}
\label{fig9}
\end{figure*}

  Figure~\ref{fig9} shows the EoS (a,c), $p(y)$, for small
        [$\kappa=1$, (a)] and large [$\kappa=10$, (c)] 
            gravitational fields,
            and the same $p(y(r))$ (b,d) but with $y=y(r)$ for the
            vdW-Skyrme
            interaction, 
  Eq.~(32)$^\ast$. Again, we compare two characteristic
  cases, small ($M=1.4 M_{\odot}$, red solid) and relatively large
  ($M=2.0 M_{\odot}$, blue frequent dashed) NS
  masses.
  The EoS pressures $p=p(y)$,
  Fig.~\ref{fig9}(a,c), are increasing functions of the density $y$
  while the pressures $p=p(y(r))$, (b,d), are decreasing as functions of
  the radial variable $r$ because of the leptodermic properties of the
  particle number density $y(r)$ for the vdW-Skyrme forces,
  Eq.~(32)$^\ast$.
  The surface ($S$) components, $p^{}_{1}$,
  Eq.~(\ref{p1cal}), are shown by black solid ($M=1.4 M_{\odot}$) and
  green long dashed ($M=2.0 M_{\odot}$) lines,
  respectively. The surface effects become much more significant with
  increasing incompressibility parameter $\kappa$, 
      that corresponds to
  a stronger gravitation field. 
      These results for the pressure are basically in agreement with those
  presented in Fig.~\ref{fig4} for the macroscopic EoS. However, the surface
  correction $p^{}_1$ leads to a more wide coordinate dependence
  of the total pressure $p$ near the ES (Fig.~\ref{fig9} (b,d)). 
   As mentioned above, the NS masses $M$ and effective radii $R$
  correspond largely to very accurate experimental data;
  see Refs.~\cite{GR21,CC20,TR21,OL20}).
      Notice also that the mass
  dependence is also essential for all results presented in Fig.~\ref{fig9} for
  the volume and surface contributions especially for a large
  gravitational field.
  We should emphasize that our results for the volume ($p^{}_{V}$)
  contribution to the total
  pressure $p$ are in good agreement with the numerical calculations
  based on the TOV equations discussed in Ref.~\cite{MM24}.

     \section{Conclusions}
\l{concl}

     The effective surface approximation based on the leptodermic expansion
     over a small parameter - a ratio of the crust thickness $a$
     to the effective radius $R$ of the system - is extended for
     the description
     of neutron star properties.
   The neutron star was considered  as 
   a finite dense and perfect liquid (including amorphous solid)
       system at
 the equilibrium.   The mean gravitational field
  $\Phi(\rho)$ was taken into account 
    in the simplest form as expansion over powers of
 differences $\rho-\overline{\rho} $, where
        $\overline{\rho}$ is the saturation density,
 up to second order in terms of the 
 particle separation
 energy
 and incompressibility. Taking into account the gradient terms of the energy
 density in a rather
 general form we
 analytically obtained the leading (over a small parameter $a/R \ll 1$)
 particle number 
 density $\rho$.
 The density $\rho$ was derived as function of the normal-to-ES coordinate
 $\xi$ of the orthogonal local nonlinear-coordinate system
 $\xi, \eta $ 
 through the effective surface. With its help, one finds
 the macroscopic equation of state (EoS)
 for the pressure $P=P(\rho)$.
 The TOV equations were re-derived 
    within 
   the
  same ESA approximation
  taking into account the leptodermic surface
  corrections.  Our results for the leading order volume pressure
  components    are in good agreement with those
  of Tolman \cite{RT87} 
      and numerical \cite{MM24} calculations.
 We  
  obtained the first order
      leptodermic surface corrections
      owing to the Schwartzschild metric, and
      corresponding modified TOV equations,
  and have solved them analytically within the same order.
  These first order corrections,
  being
  of the order of $a/R$, are also proportional
  to the dimensionless incompressibility
  depending essentially
  on the gravitational interaction constant,
  along with the nuclear interaction part.
  The surface pressure corrections are significantly increased
  with increasing NS mass, and even more pronouncely, with the gravitational
  contribution to the total incompressibility.
 
  As perspectives, one can generalize  our analytical approach to the
    the many-component and rotating systems. It is especially
  interesting to extend
  this approach to take into account  the symmetry energy
  of the isotopically asymmetric systems, 
  and the collective NS rotations within
      the Kerr metric approach at small rotational frequencies and
      deformations.
  As this approach was formulated in the local nonlinear system of
  coordinates
  $\xi, \eta$ 
  for deformed 
  shapes of the NS surface,
  one can apply our method to the pair of
  rotating neutron star pulsars.

\acknowledgments

The authors greatly acknowledge  C.A. Chin, V.Z.\ Goldberg, S.N.\ Fedotkin,
J.\ Holt, C.M.\ Ko, E.I.\ Koshchiy, J.B.\ Natowitz, A.I.\ Sanzhur, and
G.V.\ Rogachev 
for many creative and useful discussions.
A.\ Bonasera, T.\ Depastas, and A.G.\ Magner
    are partially supported by the US Department of
    Energy under Grant no. DE-FG03-93ER-40773.

\vspace{-0.6cm}
\appendix

\setcounter{equation}{0}
\renewcommand{\theequation}{A\arabic{equation}}
\section{Derivations for $\lambda_1$}
\l{appA}

Substituting Eqs.~(\ref{enerden1}) and (\ref{01})
into the second equation in
Eq.~(\ref{lamdanueq}), 
one has
at zero order leptodermic parameter  ($G=c=1$),
\be\l{eq2lam0}
8\pi \epsi_0 = e^{-\lambda^{}_0}\left(\frac{1}{r}
\frac{{\rm d}\lambda_0}{{\rm d}r}
-\frac{1}{r^2}\right) +
\frac{1}{r^2}~.
\ee
Then, using this zero order equation,
from Eq.~(\ref{lamdanueq}), one obtains in the
linear order over $\lambda^{}_1$,
\be\l{eq2lam1}
8\pi \epsi_1=\frac{1}{r}e^{-\lambda^{}_0}
  \frac{{\rm d} \lambda^{}_1}{{\rm d} r}-
  \lambda^{}_1\left(8 \pi \epsi_0-\frac{1}{r^2}\right).
\ee
This equation has the standard form
    with respect to the surface correction
$\lambda^{}_1$ as the first order linear
differential equation,
\be\l{eq2lam1-1}
\frac{{\rm d} \lambda^{}_1}{{\rm d} r} +\alpha  \lambda^{}_1=\alpha^{}_1~,
\ee
where
\bea\l{bet}
&\alpha(r)=r e^{\lambda^{}_0}\left(- 8\pi\epsi_0 + \frac{1}{r^2}\right)
\nonumber\\
&=\frac{1}{r(1-\tilde{r}^2)} - \frac{8\pi\epsi_0 r}{1-\tilde{r}^2},
\eea
$\tilde{r}=r/R_{\rm S}$, and
\bea\l{gam}
&\alpha^{}_1=\frac{8\pi r\epsi_1}{1-\tilde{r}^2}\approx
\frac{8 \pi r \mathcal{C}}{1-\tilde{r}^2}
\left(\frac{{\rm d} \rho}{{\rm d} r}\right)^2\nonumber\\
&\approx \frac{8 \pi r \varepsilon^{}_G}{1-\tilde{r}^2}~.
\eea
Here,
\be\l{e1}
\epsi_1\approx
\mathcal{C} \left(\frac{{\rm d} \rho}{{\rm d} r}\right)^2\approx
\varepsilon^{}_G ~;
\ee
see Eq.~(\ref{lam0}) for $\lambda^{}_0$ and Eq.~(\ref{01}) for $\epsi_1$ at
the leading order of the leptodermic expansion for $r < R_{\rm S}$
($\tilde{r} < 1$),
as well as Eq.~(24)$^\ast$
for the last approximation. According to Eq.~(\ref{epsKG})
for $\varepsilon^{}_G$, one has $\varepsilon^{}_G\propto y(1-y)^2$, i.e.,
$\varepsilon^{}_G$ has a sharp maximum of the width $a$
near the boundary ES (see
Fig.~\ref{fig1}). 

Using the method of an arbitrary constant variation, one obtains from 
Eq.~(\ref{eq2lam1}) the solution for $\lambda^{}_1$ in the form:
\be\l{sollam1}
\lambda^{}_1=C_1^{(\lambda)}(r)\exp\left(-\int {\rm d} r \alpha(r)\right),
\ee
where
\be\l{C}
 C_1^{(\lambda)}(r)=\int {\rm d} r \alpha^{}_1(r)
\exp\left(\int {\rm d} r\alpha(r)\right).
\ee
For the integrals in exponent of
Eq.~(\ref{sollam1}) and (\ref{C}), one obtains
\be\l{int}
\!\int \!{\rm d} r \alpha(r)
\!=\! \ln\left(\frac{\tilde{r}}{\sqrt{1\!-\!\tilde{r}^2}}\right)
\!+\!8\pi \epsi_0 R^2_{\rm S}\ln\sqrt{1\!-\!\tilde{r}^2},
\ee
i.e.,
\be\l{expint}
\exp\left(\int {\rm d} r \alpha(r)\right)=
\tilde{r}\left(1-\tilde{r}^2\right)^{-1/2+4 \pi \epsi_0R_{\rm S}^2}.
\ee
Therefore, for $C^{(\lambda)}_1(r)$, Eq.~(\ref{C}), one has
\be\l{C1}
\!C_1^{(\lambda)}(r)\!=\!8\pi\int \!r {\rm d} r \epsi_1(r)
\!\left(1\!-\!\tilde{r}^2\right)^{4\pi \epsi_0R^2_{\rm S}-3/2}.
\ee

Taking into account that $\alpha^{}_1\propto \epsi_1\propto
({\rm d} \rho/{\rm d} r)^2$,
Eq.~(\ref{gam}), at leading order of the leptodermic expansion, one can
take the smooth 
coefficient in the integrand of
Eq.~(\ref{C1}) off the integral at the ES, $r=R$. 
    The reason is that
the derivative
square, $(\partial \rho/\partial r)^2$, is a bell-like
function of $r$ with a relatively
small width of the order of the
crust thickness $a$. 
    We assume here that $r$ is not close to the
Schwarzschild radius $R_{\rm S}$, where the integrand in Eq.~(\ref{C1})
is singular.
 In line of Ref.~\cite{MM24}, the remaining integration
within the same precision
can be taken analytically. Finally, 
    for $C^{(\lambda)}_1(r)$ one
finds ($r < R_{\rm S}$)
\be\l{C1fin}
C_1^{(\lambda)}(r)=8\pi R \left(1-\tilde{R}^2\right)^{4\pi\epsi_0R_{\rm S}^2-3/2}
\mathcal{D}_1~,
\ee
    where $\mathcal{D}_1$ is given by Eq.~(\ref{e1intdef}); see also
    Eq.~(\ref{e1intout}) for the asymptote
    outside and Eq.~(\ref{e1intin})
[in terms of $\mathcal{Y}(y)$, Eq.~(\ref{Y})] inside of the system.
The effective radius $R$ is presented 
in units of $R_{\rm S}$. From Eqs.~(\ref{C1fin}) and
(\ref{expint}), one arrives at the final expression (\ref{sollam1final})
for $\lambda^{}_1$.

Using Eqs.~(\ref{enerden1}) for $\epsi$ and
(\ref{01}) for $P$ and $\lambda$, and Eq.~(\ref{eq2lam0}) for
$P_0$ at the zero order of leptodermic parameter
$a/R$, one can derive the
linearized equation (\ref{eqP1}) for $P_1$
beyond the standard TOV equations
(\ref{eq2lam0}) and (\ref{toveq}) for $P_0$.
Equation~(\ref{eqP1}) for the leptodermic pressure
correction $P_1$ has the same standard form as Eq.~(\ref{eq2lam1})
for $\lambda^{}_1$. Using also
the same method of the arbitrary constant variations,
one obtains the solution in terms of the quadratures, Eqs.~(\ref{solP1})
for $P_1$ and (\ref{solCp}) for $C_1^{(P)}$.
Substituting the function $\beta(r)$, Eq.~(\ref{qRs}),
into the integrals in Eqs.~(\ref{solP1}) and
(\ref{solCp}), one finds Eqs.~(\ref{qintRs}) for $\int {\rm d} r \beta(r)$
and (\ref{solCp1}) for $C_1^{(P)}$.
As shown in section \ref{mtov},
within the first order over $a/R$, one has to neglect $\lambda^{}_1$
which gives the second order leptodermic correction to the
pressure correction
$P_1$. Finally, with Eq.~(\ref{solCpRs1}) for $C^{(P)}_1$,
for the leptodermic correction
$P_1$, one arrives at Eq.~(\ref{solP1fin}).

\vspace{-0.2cm}


\begin{thebibliography}{99}

  \bibitem{wilets} L. Wilets, Phys.Rev. {\bf 101}, 1805 (1956); 
  Rev. Mod. Phys. {\bf 30}, 542 (1958).

\bibitem{strtyap} V.M. Strutinsky, and A.S. Tyapin, Exp. Theor. Phys. (USSR) 
{\bf 18}, 664 (1964).

\bibitem{tyapin} A.S. Tyapin, Sov. Journ. Nucl. Phys. {\bf 11}, 401 (1970),
  {\bf 13}, 32 (1971), {\bf 14}, 50 (1972).

\bibitem{strmagbr} V.M. Strutinsky, A.G. Magner, and M. Brack, Z. Phys. A
{\bf 319}, 205 (1984).

\bibitem{strmagden} V.M. Strutinsky, A.G. Magner, and V. Yu. Denisov, Z. Phys.,
  {\bf 322}, 149 (1985).

\bibitem{MS09} A.G.\ Magner, A.I.\ Sanzhur, and A.M.\ Gzhebinsky, Int. J.
  Mod. Phys. E {\bf 92}, 064311 (2009).

\bibitem{BM13}J.P.\ Blocki, A.G.\ Magner, P.\ Ring, and A.A.\ Vlasenko,
  Phys. Rev. C {\bf 87}, 044304 (2013).

  \bibitem{BM15} J.P.\ Blocki, A.G.\ Magner, and P.\ Ring,
    Phys. Rev. C {\bf 92}, 064311 (2015).

  \bibitem{MM24} A.G.\ Magner, S.P.\ Maydanyuk,
    A.\ Bonasera, H.\ Zheng, T.\ Depastas, A.I.\ Levon, and U.V.\ Grygoriev,
    Int. Journ. Mod. Phys. E {\bf 33}, 2450043 (2024);
    arXiv:2403.01445v3 [nucl-th] (2024).
   
\bibitem{RT39} R.C.\ Tolman, Phys. Rev. C{\bf 55}, 364 (1939).

      \bibitem{OV39} J.R.\ Oppenheimer and G.M.\ Volkoff,
      Phys. Rev. C{\bf 55}, 374 (1939).

  \bibitem{RT87} R.C.\ Tolman, {\it Relativity, Thermodynamics, and Cosmology},
    (Dover Publications, New York, 1987; Oxford, the University Press,
    1934, 1946, 1949, 1987).

       \bibitem{LLv2} L.D.\ Landau and E.M.\ Lifshitz, {\it
      Theoretical Physics, v.2,
      The Classical Theory of Fields} (Elsevier, Amsterdam,1951-1971;
      FIZMATGIZ, Moscow, 2003).

      \bibitem{LLv6} L.D.\ Landau and E.M.\ Lifshitz,
  {\it Theoretical Physics, v.6,
    Fluid Mechanics} (Elsevier, Amsterdam, 1959-1987).

\bibitem{RW82}
      J.S.\ Rowlinson and B.\ Widom, {\it Molecular Theory of
  Capillarity} (Clarendon Press, Oxford, 1982).
  

   \bibitem{KS16} K.\ Schwarzschild,
     Berl. Ber., p. 424 (1916).

  \bibitem{WF88} R.B.\ Wiringa, V.\ Fiks, A.\ Fabrocini,
    Phys. Rev. C {\bf 38},
    1010 (1988).

    \bibitem{ST04}  S.L.\ Shapiro, S.A.\ Teukolsky, {\it
  Black holes, white dwarfs,
     and neutron stars: The physics of compact objects} (Wiley-VCH Verlag
     GmbH\& Co.KGaA, Weinheim, 2004).

\bibitem{CB97} E.\ Chabanat, P.\ Bonche, P.Haensel, J.\ Meyer, and
  R.\ Shaeffer, Nucl. Phys. A {\bf 627}, 710 (1997).

\bibitem{SH06} I. Sagert, M. Hempel, C. Greiner,
    and J. Schaffner-Bielich,
    Eur, J. Phys. {\bf 27}, 577 (2006).

    \bibitem{HPY07}  P.\ Haensel, A.Y.\ Potehin, D.G.\ Yakovlev,
   Astrophysics and space science library, vol. {\bf 326}, {\it Neutron Stars 1.
     Equation of State and Structure} (Springer, New York, 2007). 

\bibitem{Ko08} Bao-An Li, Lie-Wen Chen, and Che Ming Ko, Phys. Rep.,
  {\bf 464}, 113 (2008).

\bibitem{PFCPS13} 
    A.Y.\ Potekhin, A. F.\ Fantina, N.\ Chamel, J.M.\ Pearson,
  and S.\ Goriely, Astronomy \& Astrophysics, {\bf 560}, A48 (2013).

\bibitem{AB14} G.\ Giuliani, H.\ Zheng, A.\ Bonasera, Prog. Part. Nucl. Phys.
  {\bf 76}, 116 (2014).

\bibitem{LH19} Y.\ Lim and J.W.\ Holt, Eur.Phys. J. A {\bf 55}, 209 (2019).

    \bibitem{SBL23} Boyang Sun, Saketh Bhattiprolu, and James M. Lattimer,
      arXiv:2311.00843v1, 2023.

    \bibitem{BC15} R.\ Belverde, F.\ Cipolletta, C.\ Cherubini,
  S.M.\ de Carvalho, S.\ Filippi, R.\ Negreiros, J. P. Pereira,
  J.A.\ Rueda, and R.\ Ruffini, {\it Physics and Astrophysics of
    Neutron Stars,} The second ICRANet C\'esar Lattes Meeting, AIP Conf. Proc.
  1693,030001-1-030001-19; doi;10.1063/1.4937184 (AIP Publishing LLC
  978-0-7354-5/\$30.00), p.030001-1 (2015).

    \bibitem{BBP71} G.~Baym, H.~Bethe, and C.J.~Pethick,
    Nucl. Phys. A {\bf 175} (1971) 225. 

  \bibitem{FCPG13} A.F.\ Fantina, N.\ Chamel,
  J.M.\ Pearson, and S. Goriely, Astronomy \& Astrophysics,
  {\bf 559}, A128 (2013).

     \bibitem{LJ23} 
      J.M.\ Lattimer, J. Phys., Conf. Ser., {\bf 2536}, 012009
     (2023).

     \bibitem{LP01} 
         J.M.\ Lattimer and M.\ Prakash, The astrophysical Journal,
       {\bf 550}, 426 (2001).

     \bibitem{HP01PRL} 
         C.J.\ Horowitz and J.\ Piekarewicz, Phys. Rev. Lett.,
       {\bf 86}, 5647 (2001).

   \bibitem{HP01PRC}  C.J.\ Horowitz and J.\ Piekarewicz, Phys. Rev. C,
       {\bf 64}, 062802(R) (2001).
     
\bibitem{PH00Poland}  P.Haensel, Neutron Star Crusts,
       N. Copernicus Astronomical Center, Polish Academy of Sciences,
       Bartycka 18, PL-00-716 Warszawa, Poland, 2000.

     \bibitem{CH08} N.\ Chamel and P.\ Haensel, Liv. Rev. Relativ. {\bf 11},
  10 (2008).    

   \bibitem{ABCG09} Z.\ Arzoumanian,
       S.\ Bogdanov, J.\ Cordes, K.\ Gendreau,
     D.\ Lai, J.\ Lattimer,
B.\ Link, A.\ Lommen, C.\ Miller, P.\ Ray, R.\ Rutledge, T.\ Strohmayer,
C.\ Wilson-Hodge, and K.\ Wood, arXiv:0902.3264~~, 2009; 	
https://doi.org/10.48550/arXiv.0902.3264.

   \bibitem{FG23} A.F.\ Fantina, and F.\ Gulminelli,
  J. Phys., Conf. Ser. {\bf 2586}, 012112 (2023);
  doi:10.1088/1742-6596/2586/1/012112~.

\bibitem{DFG23} H.\ Dinh, A.F.\ Fantina, and F.\ Gulminelli,
  Eur. Phys. J. A {\bf 59}, 292(2023);
  https://doi.org/10.1140/epja/s10050-023-01199-x~. 

\bibitem{Pe23} N. N.\ Shchechilin, N.\ Chamel, J. M.\ Pearson,
  Phys. Rev. C {\bf 108}, 025805 (2023).

\bibitem{Pe24} N.\ Chamel, J.M.\ Pearson, and N.N.\ Shchechilin,
  arXiv:2410.01997v2 [nucl-th] (2024).

\bibitem{XV24}  
    D.\ Kobyakov and X. Vi\~{n}as, arXiv:2411.17303v1 [nucl-th]
    (2024).

 \bibitem{vdW} J.S.\ Rowlinson, Journ. Stat. Phys. {\bf 20}, 197 (1979). 
  
\bibitem{bete} H.\ Bethe, Annu. Rev. Nucl. Sci., {\bf 21}, 93 (1971). 

\bibitem{brbhad}
M.\ Brack and R.\ K.\ Bhaduri, {\it Semiclassical Physics.
  Frontiers in Physics} No. {\bf 96}, 2nd ed.
(Westview Press, Boulder CO, USA, 2003).


  M.\ Brack and R. K.\ Bhaduri, \textit{Semiclassical Physics}
(Addison-Wesley, Reading MA) 1997; 2nd edition (Westview Press, Boulder) 2003.

\bibitem{brguehak} M.\ Brack, G.\ Guet, H.-B.\ H{\aa}kansson, 
  Phys. Rep. {\bf 123}, 275 (1985).

\bibitem{bormot} A.\ Bohr and B.\ Mottelson, \textit{Nuclear structure} (W.
  A. Benjamin, New York, 1975), Vol. II.

\bibitem{magstr} A. G.\ Magner, V. M.\ Strutinsky, Z. Phys. A \textbf{324}, 633
  (1985).

\bibitem{Ti52}  A.N.\ Tikhonov,
  {\it Systems of differential equations containing
  a small parameter in front of the derivatives},
  Mat. Sb. , 31 : 3 (1952) pp. 575-586 (In Russian).

\bibitem{vauthbrink} D.\ Vautherin, D.\ Brink, Phys. Rev. C {\bf 5} 626 (1972). 

\bibitem{skyrme} T.H.R.\ Skyrme, Philos. Mag., {\bf 1} 8th ser., 1043 (1956).

  \bibitem{barjac} R.C.\ Barrett, D.F.\ Jacson, 
    \textit{Nuclear sizes and structure} Oxford, Clarendon Press 1977.

  \bibitem{ringshuk} P.\ Ring, P.\ Schuck,
    \textit{The nuclear many-body problem,} 
Berlin, Heidelberg, New York, Springer-Verlag 1980.
 

\bibitem{blaizot} J.P.\ Blaizot, Phys. Rep. {\bf 64}, 172 (1980).

 \bibitem{krivin} H.\ Krivine, J.\ Treiner, and O.\ Bohigas, Nucl. Phys. A, 
   {\bf 336}, 155, (1980).

\bibitem{CB98} E.\ Chabanat, P.\ Bonche, P.Haensel, J.\ Meyer, and
    R.\ Shaeffer, Nucl. Phys. A {\bf 635}, 231 (1998).

\bibitem{gramvoros} B. Gramaticos, A. Voros, Ann.Phys., {\bf 123}, 359 (1979);
{\bf 129}, 153 (1980.)
  

\bibitem{kolsan} V.M.\ Kolomietz and A.I.\ Sanzhur,
  Eur. Phys. J. A {\bf 38}, 345–354 (2008).

\bibitem{KS18} V.M.\ Kolomietz, A.I.\ Sanzhur, and S.\ Shlomo,
  Phys. Rev. C {\bf 97}, 064302 (2018).

\bibitem{KS20} V.M.\ Kolomietz and S.\ Shlomo, Mean Field Theory,
  World Scientific, Singapore (2020).

  \bibitem{myswann69} W.D.\ Myers and W.J.\ Swiatecki,
  Ann. Phys. (NY) {\bf 55}, 395 (1969);  {\bf 84}, 186 (1974).


\bibitem{myswiat85} W.D.\ Myers, W.J.\ Swiatecki, and C.S.\ Wang,
  Nucl. Phys. A {\bf 436}, 185 (1985).

  \bibitem{danielewicz2} P.\ Danielewicz and J.\ Lee, Int. J. Mod. Phys.
    E {\bf 18}, 892 (2009).

    \bibitem{vinas1} M.\ Centelles, X.\ Roca-Maza, X.\ Vi\~{n}as, and M.\ Warda,
Phys. Rev. Lett., {\bf 102}, 122502 (2009).

%
\bibitem{vinas2} M.\ Warda, X.\ Vi\~{n}as, X.\ Roca-Maza, and M.\ Centelles,
Phys. Rev. C {\bf 80}, 024316 (2009); {\bf 81}, 054309 (2010); 
{\bf 82}, 054314 (2010); arXiv:0906.0932 [nucl-th] (2009).

%
\bibitem{vinas4}  X.\ Roca-Maza,  M.\ Centelles, X.\ Vi\~{n}as, and M.\ Warda,
  Phys. Rev. Lett. {\bf 106}, 252501 (2011).

\bibitem{vinas5}   X.\ Vi\~{n}as,  M.\ Centelles, X.\ Roca-Maza,
  and M.\ Warda,
    Eur. Phys. J. A  {\bf 50}, 27 (2014).

   \bibitem{Pi09} J.\ Piekarewicz and M.\ Centelles, Phys.~Rev.~ C {\bf 79},
  054311 (2009). 

\bibitem{NVR11} T.\ Niksic, D.\ Vretenar, and P.\ Ring,
  Prog. Part. Nucl. Phys., {\bf 66}, 519 (2011).


\bibitem{Pi21} B.D.\ Reed, F.J.\ Fattoyev,
  C.J.\ Horowitz, and J.\ Piekarewicz, Phys.~Rev.~ Lett.~ {\bf 126},
  172503 (2021). 


\bibitem{GR21} G. Raaijmaket et al.,  The Astrophysical Journ. Lett.
  {\bf 918}: L29 (2021),  1 (13 pp).
  
\bibitem{CC20} C.D. Capano et al.,
    Nature Astronomy {\bf 4}, (2020) 625.


\bibitem{TR21}
  T.E. Riley et al.,
    The Astrophysical Journ. Lett. {\bf 918}: L27 (2021), 1 (30 pp).

\bibitem{OL20} O.\ Lorenco, M.\ Bhuyan, C.H.\ Lenzi, M.\ Dutra,
  C.\ Gonzalez-Boquera, M.\ Centelles, X.\ Vi\~{n}as, Phys. Lett.
  B {\bf 803}, (2020) 135306.  

\end{thebibliography}
\end{document}